\documentclass[ reprint,
superscriptaddress,
bibnotes,
 amsmath,amssymb,
 aps,
pra,
]{revtex4-2}
\usepackage{lipsum}
\usepackage{graphicx}
\usepackage{soul} 
\usepackage{dcolumn}
\usepackage{bm}
\usepackage[caption=false]{subfig} 
\usepackage{cancel}
\usepackage{comment}
\usepackage{mathrsfs}
\usepackage{orcidlink}
\usepackage{tikz}
\usepackage[dvipsnames]{xcolor}
\usepackage{hyperref}

\hypersetup{
    colorlinks=true,
    linkcolor=Blue,
    citecolor=Blue,
    filecolor=Blue,      
    urlcolor=Blue,
    }

\begin{document}

\preprint{APS/123-QED}

\title{Experimental signatures of bosonic pairing in a two-component Bose gas}
\author{Omar Almonajed\,\orcidlink{0009-0001-2816-7959}}
 \affiliation{%
Department of Physics, Bilkent University, Ankara, 06800, T\"urkiye
}%

\author{M. Iskin\,\orcidlink{0000-0003-0704-1318}}
 \affiliation{%
Department of Physics, Ko\c{c} University, Rumelifeneri Yolu, 
34450 Sar\i yer, Istanbul, T\"urkiye
}%

\author{M. \"O. Oktel$^1$\,\orcidlink{0000-0001-8921-8388}}

\date{\today}

\begin{abstract}

Two-component bosonic droplets are commonly described within Bogoliubov 
theory, where beyond-mean-field quantum fluctuations stabilize the system 
against mean-field collapse. In the interaction regime where droplets form, 
however, the Bogoliubov excitation spectrum contains an imaginary branch 
associated with the underlying instability, which is typically omitted when
evaluating the beyond-mean-field energy. Bosonic pairing theory provides 
an alternative description with a fully real excitation spectrum. 
In this work, we reformulate bosonic pairing theory within an operator 
formalism, making its underlying approximations transparent, and compare 
its predictions with those of Bogoliubov theory for a homonuclear binary 
Bose mixture across the crossover from a weakly interacting gas to the 
droplet regime. Working at fixed particle density, we determine the variational 
parameters of bosonic pairing theory self-consistently and focus on the 
regime in which both theories possess real excitation spectra, allowing 
a direct comparison. We find that bosonic pairing theory yields a lower 
grand potential than Bogoliubov theory and predicts qualitatively distinct 
correlation signatures, including enhanced long-wavelength interspecies 
momentum-space correlations and a finite density static structure factor 
at experimentally accessible low momenta. These differences persist over 
a broad range of interaction strengths and suggest that correlation and 
structure-factor measurements can provide direct experimental tests of 
bosonic pairing across the gas-to-droplet crossover.

\end{abstract}

\maketitle

\section{Introduction}
\label{sec:intro}

Ultradilute quantum droplets~\cite{petrov_liquid_2018} are self-bound 
quantum phases of matter characterized by a finite equilibrium density. 
Unlike trapped atomic gases, which require external confinement, 
droplets are stabilized by an intrinsic balance between attractive 
mean-field interactions and repulsive beyond-mean-field quantum
fluctuations~\cite{petrov_quantum_2015}. They therefore provide 
a particularly clean platform for exploring correlation effects 
beyond the mean-field paradigm~\cite{skov_observation_2021}. 
Following their theoretical prediction in Bose-Bose
mixtures~\cite{petrov_quantum_2015}, 
quantum droplets have been experimentally realized in a variety of systems, 
including dipolar gases~\cite{schmitt_self-bound_2016,
ferrier-barbut_observation_2016, chomaz_quantum-fluctuation-driven_2016, zhang2026observation} 
and Bose-Bose mixtures with short-range interactions in both homonuclear
\cite{cabrera_quantum_2018, cheiney_bright_2018, semeghini_self-bound_2018} 
and heteronuclear configurations~\cite{derrico_observation_2019}. 
Related self-bound states have also been theoretically proposed in 
Bose-Fermi mixtures
\cite{cui_spin-orbit-coupling-induced_2018, rakshit_quantum_2019, rakshit_self-bound_2019},
lower-dimensional Bose mixtures~\cite{petrov2016, spada24}, 
and spinor Bose gases~\cite{yogurt2022, yogurt2023}.

For a two-component Bose gas with repulsive intraspecies and attractive
interspecies interactions, increasing the magnitude of the interspecies
attraction suppresses the leading mean-field energy and drives the
system toward an expected collapse. However, higher-order quantum
fluctuation effects provide stability and result in the self-bound droplet state.
\cite{petrov_quantum_2015}. Standard Bogoliubov theory captures this mechanism
through the Lee-Huang-Yang (LHY) correction, whose repulsive energy
counteracts the attractive mean-field contribution and produces a
self-bound equilibrium state with zero pressure. However, the
quadratic excitation spectrum simultaneously develops a soft mode that
becomes unstable in the droplet regime. In Petrov's
treatment~\cite{petrov_quantum_2015}, the contribution of this unstable branch
is omitted when evaluating the fluctuation energy.

Bosonic pairing theory provides an alternative description that resolves this
tension between thermodynamic stability and spectral instability
\cite{hu_consistent_2020,hu_microscopic_2020}. By introducing an ad hoc BCS-like
pairing field, it preserves the same physical stabilization mechanism,
namely the competition between attractive mean-field and repulsive
fluctuation energies, while replacing the softening branch with a gapped
collective mode in the droplet regime. In the formulation of Hu and Liu,
the paired state coexists with a single-particle condensate and therefore
corresponds to what Nozi\`eres and Saint James termed a \emph{mixture state},
rather than a purely paired condensate~\cite{nozieres1982particle}.
More generally, bosonic pairing forms part of a broader body of work aimed
at developing spectrally stable descriptions of attractive Bose systems
through self-consistent treatments of pairing and quantum fluctuations
\cite{nozieres1982particle, jeon2002, koetsier2009, yu2010, iskin2020,
wang2025, hohenadler2010, ota2020, gu2020, pawel2021, zin2022, rakhimov2026}.
The Hu-Liu formulation is one of the latest developments in this
ongoing effort to formulate a microscopic theory that simultaneously
captures droplet stabilization and a quasiparticle spectrum free of
dynamical instabilities~\cite{hu_consistent_2020,hu_microscopic_2020}.
Unlike Bogoliubov theory, the pairing description remains nonsingular at
the mean-field collapse point and can therefore be continued smoothly
into the weakly attractive gas phase. If it provides an accurate description
of droplets, one may therefore expect it to remain a useful approximation
on the gas side of the crossover as well, motivating the comparison carried
out in this work.

The existing formulation of bosonic pairing theory is based on a 
functional-integral approach in which a spatially uniform pairing field 
is introduced and determined self-consistently. As a first step, we rederive 
the theory using an operator formalism. This complementary derivation 
provides a more transparent physical interpretation of the underlying 
approximations by separating the introduction of the pairing field 
from the treatment of density fluctuations. The resulting formalism is 
equivalent to the original theory while making its assumptions more explicit 
and providing a natural framework for comparison with Bogoliubov theory.

Previous studies of bosonic pairing theory have focused primarily on 
equilibrium properties of self-bound droplets, such as the equilibrium 
density, droplet stability, and density 
locking~\cite{hu_breakdown_2025, hu_microscopic_2020}. Consequently, 
relatively little is known about how the theory modifies experimentally 
accessible observables throughout the attractive-interaction regime, 
particularly in the gas phase where no droplet is present. 
Understanding these differences is especially important because the 
two theories predict broadly similar equilibrium thermodynamics despite 
possessing qualitatively different quasiparticle spectra. 
Moreover, measurements in the gas phase are experimentally advantageous, 
as the lower densities substantially suppress three-body losses, 
providing a cleaner setting for distinguishing between the predictions 
of the two theories.
Recent experimental advances now make such comparisons feasible.
In-situ density-correlation measurements~\cite{yao_measuring_2025, de_jongh_quantum_2025},
momentum-space measurements of two-body and higher-order correlations
\cite{tenart_observation_2021, bureik_suppression_2025}, and structure-factor 
measurement techniques developed for ultracold-atom systems
\cite{stenger_bragg_1999, shammass_phonon_2012} provide direct access 
to many-body correlations beyond equilibrium thermodynamic quantities.

In this paper, we present a systematic comparison between Bogoliubov 
theory and bosonic pairing theory for equal-mass, equal-interaction, 
equal-density binary Bose mixtures at fixed particle density across 
the entire interspecies-attractive regime, including the crossover 
into the droplet phase. We focus on this symmetric case to provide 
the clearest possible understanding of the underlying physics, although 
the formalism can be extended straightforwardly to more general mixtures. 
We show that bosonic pairing theory consistently yields a lower 
grand potential than Bogoliubov theory while maintaining a fully 
real quasiparticle spectrum. We then investigate interspecies two-body 
correlations in both momentum and position space, together with 
branch-resolved and species-resolved static structure factors. 
Our central result is that bosonic pairing theory predicts qualitative 
correlation signatures absent from conventional Bogoliubov theory despite 
the two approaches exhibiting broadly similar equilibrium thermodynamics. 
Most notably, it produces a zero-temperature bunching feature in momentum-space 
interspecies correlations extending to momenta of order the healing 
length, a finite low-momentum interspecies structure factor, and 
systematically modified short-range interspecies correlations in real space.

The remainder of this paper is organized as follows.
Section~\ref{sec:bog} revisits Bogoliubov theory for homonuclear,
equal-density binary Bose mixture and establishes the quadratic
framework used throughout the paper. Section~\ref{sec:pairing}
introduces the central approximation underlying bosonic pairing theory
and compares the resulting grand potentials over a range of
interspecies interaction strengths at fixed density.
Sections~\ref{sec:corr} and \ref{sec:sf} present the momentum-space and
real-space two-body correlations, together with the branch-resolved and
species-resolved static structure factors, for both theories.

\section{Bogoliubov Theory} 
\label{sec:bog}

To describe a uniform two-component Bose gas in three dimensions,
we consider the Hamiltonian
$
\mathcal{H}=\mathcal{H}_0+\mathcal{H}_\mathrm{int},
$
where $\mathcal H_0$ accounts for the single-particle dynamics
and $\mathcal H_\mathrm{int}$ describes short-range density-density
interactions:
\begin{align}
\mathcal{H}_{0}
&=
\sum_{i\mathbf{k}}
(\epsilon_{\mathbf{k}}-\mu)
c_{i\mathbf{k}}^{\dagger}c_{i\mathbf{k}},
\nonumber\\
\mathcal{H}_\mathrm{int} &=
\sum_{ij} \frac{g_{ij}}{2V}
\sum_{\mathbf{k}\mathbf{k}'\mathbf{q}}
c_{i,\mathbf{k}+\mathbf{q}}^{\dagger}
c_{j,\mathbf{k}'-\mathbf{q}}^{\dagger}
c_{j \mathbf{k}'} c_{i \mathbf{k}}.
\label{eq:generic-exactH}
\end{align}
Here, $c_{i\mathbf{k}}^{\dagger}$ creates a boson of species
$i=(1,2)$ with momentum $\hbar \mathbf{k}$,
$\epsilon_{\mathbf{k}}=\hbar^2k^2/(2m)$ is the single-particle
dispersion, $\mu$ is the chemical potential, and $V$ is the
system volume. The interaction strengths are related to the
scattering lengths through
$
g_{ij}=4\pi\hbar^2a_{ij}/m.
$
Throughout this work, we focus on a symmetric mixture with
equal masses $m_1=m_2=m$, equal densities
$n_1=n_2=n=N/V$, equal intraspecies interactions
$g_{11}=g_{22}=g > 0$, and tunable interspecies interaction
$g_{12}=g_{21} < 0$. While all calculations can be extended to the full parameter regime, 
this restriction allows us to isolate the
essential differences between Bogoliubov theory and bosonic
pairing theory.

Bogoliubov theory describes the system as a weakly-interacting
condensate dressed by Gaussian quantum fluctuations. The
central assumption is that the zero-momentum mode is
macroscopically occupied, allowing the corresponding operators
to be replaced by classical fields,
$c_{i\mathbf 0}\approx\sqrt{N_0}$,
where $N_0$ is the condensate population of each species.
Expanding the Hamiltonian in powers of the fluctuation operators
$c_{i, \mathbf {k \ne 0}}$ then yields
$
\mathcal H_{\mathrm{Bog}}
=
\mathcal E_{\mathrm{mf}}
+
\mathcal H_2,
$
where $\mathcal E_{\mathrm{mf}}$ is the condensate contribution
and $\mathcal H_2$ describes quadratic fluctuations about the
mean-field state. The mean-field energy density is
\begin{equation} \label{eq:Bog-MF-grand-potential}
\frac{\mathcal E_{\mathrm{mf}}}{V}
= (g+g_{12})n_0^2 - 2\mu n_0,
\end{equation}
where $n_0 = N_0/V$.
The fluctuation Hamiltonian takes the form
\begin{equation} \label{eq:Bog-quad-Hamiltonian}
\begin{aligned}
\mathcal H_2 &=
\sum_{i\mathbf k}'\Big[
\xi_{\mathbf k}  c_{i\mathbf k}^{\dagger}c_{i\mathbf k}
+\frac{gn_0}{2}
\big(c_{i\mathbf k}^{\dagger}c_{i,-\mathbf k}^{\dagger}
+\mathrm{H.c.}\big)
\Big]
\\
&\quad
+ g_{12}n_0 \sum_{\mathbf k}'
\big(
c_{1\mathbf k}^{\dagger}c_{2,-\mathbf k}^{\dagger}
+
c_{1\mathbf k}^{\dagger}c_{2\mathbf k}
+\mathrm{H.c.}
\big),
\end{aligned}
\end{equation}
where the primed sums $\sum_{\mathbf k}'$ denote the exclusion 
of the condensate mode $\mathbf{k = 0}$, and
$
\xi_{\mathbf{k}}
=
\epsilon_{\mathbf{k}}
-\mu
+
(2g+g_{12})n_0
$
is introduced for convenience. The first line
contains the  intraspecies Bogoliubov processes,
while the second line shows that interspecies interactions
generate both pairing and density-exchange fluctuations,
where $\mathrm{H.c.}$ denotes the Hermitian conjugate.

The Bogoliubov approximation neglects cubic and quartic fluctuation
contributions, thereby reducing the problem to a quadratic 
Hamiltonian that can be diagonalized exactly. In homogeneous systems 
such as this one, the linear terms vanish due to momentum conservation. 
 Importantly, both normal terms
$c_{i\mathbf k}^{\dagger}c_{j\mathbf k}$
and anomalous pairing terms
$c_{i\mathbf k}^{\dagger}c_{j,-\mathbf k}^{\dagger}$
appear in both the intra- and interspecies channels. As we
show below, the retention of these interspecies normal
processes constitutes one of the central distinctions between
Bogoliubov theory and bosonic pairing theory.

To determine the collective excitation spectrum, it is
convenient to cast the quadratic Hamiltonian into
Bogoliubov-de Gennes (BdG) form. Introducing
$
\mathbf{\Psi}_{\mathbf k} = (
c_{1\mathbf k}, c_{2\mathbf k},
c_{1,-\mathbf k}^{\dagger}, c_{2,-\mathbf k}^{\dagger}
)^\mathrm{T},
$
where $\mathrm{T}$ denotes matrix transpose, the
Hamiltonian can be written as~\cite{gorkov_energy_1958,nambu_quasiparticles_1960}
\begin{equation}
\label{eq:generic-BdG-form}
\mathcal H_{\mathrm{Bog}}
=
\mathcal E_{\mathrm{mf}}
-
\sum_\mathbf k'
\xi_\mathbf k
+
\frac12
\sum_{\mathbf k}'
\mathbf{\Psi}_{\mathbf k}^{\dagger}
\mathbb M_{\mathbf k}
\mathbf{\Psi}_{\mathbf k},
\end{equation}
where the first sum originates from normal ordering 
and the matrix is
\begin{equation} \label{eq:Bog-BdG-matrix}
\mathbb{M}_{\mathbf{k}} =
\begin{pmatrix}
\xi_{\mathbf{k}} & g_{12}n_{0} & gn_{0} & g_{12}n_{0} \\
g_{12}n_{0} & \xi_{\mathbf{k}} & g_{12}n_{0} & gn_{0} \\
gn_{0} & g_{12}n_{0} & \xi_{-\mathbf{k}} & g_{12}n_{0} \\
g_{12}n_{0} & gn_{0} & g_{12}n_{0} & \xi_{-\mathbf{k}}
\end{pmatrix}.
\end{equation}
The diagonal terms depend only on the magnitude of momentum, 
$\xi_\mathbf k = \xi_{-\mathbf k}$.

The symmetry of the homonuclear, equal-density mixture simplifies 
the problem considerably. Rather than working in the species basis, 
it is advantageous to introduce density and spin fluctuations through 
the canonical transformation
$
c_{1 \mathbf{k}}
=
(c_{+,\mathbf{k}}+c_{-,\mathbf{k}})/\sqrt{2},
$
and
$
c_{2 \mathbf{k}}
=
(c_{+,\mathbf{k}}-c_{-,\mathbf{k}})/\sqrt{2}.
$
This transformation separates the collective density
($s=+$) and spin ($s=-$) degrees of freedom, causing the
four-dimensional BdG problem to decouple into two
independent two-dimensional sectors.
Thus, the Hamiltonian
takes the block-diagonal form
\begin{equation}
\label{eq:BdG-generic}
\mathcal{H}_{\rm Bog}
=
\mathcal{E}_{\rm mf}
-
\frac{1}{2}
\sum_{s\mathbf{k}}^{\prime}
A_{s\mathbf k}
+
\frac{1}{2}
\sum_{s\mathbf{k}}^{\prime}
\mathbf{\Psi}_{s\mathbf{k}}^{\dagger}
\begin{pmatrix}
A_{s\mathbf k} & B_s \\
B_s & A_{s\mathbf k}
\end{pmatrix}
\mathbf{\Psi}_{s\mathbf{k}},
\end{equation}
where the two-component operator is
$
\mathbf{\Psi}_{s\mathbf k}
=
(c_{s\mathbf k},c_{s,-\mathbf k}^{\dagger})^T.
$
For the Bogoliubov theory considered in this section,
\begin{equation}
A_{s\mathbf{k}}
= \xi_\mathbf k +
sg_{12}n_{0},
\quad
B_s
=
g_s n_0,
\label{eq:Bog-AB}
\end{equation}
where $g_s=g\pm s g_{12}$, while $s=+1$ and $s=-1$
label the density and spin branches, respectively.
More generally, Eq.~\eqref{eq:BdG-generic} provides a
common set of equations for any symmetric binary Bose mixture
treated within a quadratic approximation, with the specific
theory defined through the expressions of 
$A_{s\mathbf{k}}$ and $B_s$.

Each sector can now be diagonalized independently through
the Bogoliubov transformation
$
c_{s\mathbf{k}}
=
u_{s\mathbf{k}} b_{s\mathbf{k}}
+
v_{s\mathbf{k}} b_{s,-\mathbf{k}}^{\dagger},
$
with $u_{s \mathbf k}^2 - v_{s \mathbf k}^2=1$, yielding the diagonal Hamiltonian 
\begin{equation} \label{eq:generic-final-H}
\mathcal{H}_{\mathrm{Bog}} = \mathcal E_{\mathrm{mf}}
+ \frac12
\sum_{s\mathbf k}
(E_{s\mathbf k}-A_{s\mathbf k})
+ \sum_{s \mathbf k} E_{s \mathbf k}   b_{s \mathbf k}^\dagger b_{s \mathbf k}.
\end{equation}
Here, the zero-point contributions from the coefficient $A_{s \mathbf k}$ 
and the spectrum generates the LHY correction to the grand
potential, which we denote by $\mathcal E_{\mathrm{LHY}}$ below.
The corresponding quasiparticle spectrum is
\begin{equation}
E_{s\mathbf{k}}
=
\sqrt{A_{s\mathbf{k}}^2-B_s^2}.
\label{eq:spectrumgen}
\end{equation}
Equations~\eqref{eq:BdG-generic}, \eqref{eq:generic-final-H}, 
and \eqref{eq:spectrumgen} apply to both Bogoliubov and bosonic 
pairing theories. The distinction lies solely in the coefficients 
$A_{s\mathbf{k}}$ and $B_s$.

The Bogoliubov transformation coefficients are obtained by
requiring that the quasiparticle Hamiltonian be diagonal in
terms of the new bosonic operators $b_{s\mathbf k}$. This
condition eliminates the anomalous terms and yields the transformation 
coefficients
\begin{equation}
u_{s\mathbf{k}}^{2} =
\frac{1}{2}
\Big(
\frac{A_{s\mathbf{k}}}{E_{s\mathbf{k}}}
+1
\Big),
\qquad
v_{s\mathbf{k}}^{2} =
\frac{1}{2}
\Big(
\frac{A_{s\mathbf{k}}}{E_{s\mathbf{k}}}
-1
\Big),
\label{eq:generic-usq-vsq}
\end{equation}
which quantify the particle-hole admixture of the collective
excitations. The corresponding product is
\begin{equation}
\label{eq:generic-uv}
u_{s\mathbf{k}}v_{s\mathbf{k}}
=
-\frac{B_s}{2E_{s\mathbf{k}}},
\end{equation}
where the negative-sign convention is chosen so that the
single-species limit $a_{12}\to0$ continuously recovers the
standard Bogoliubov result~\cite{fetter_quantum_1971}. These
coefficients play a central role in determining the
quantum depletion, anomalous correlations, and structure
factors discussed in later sections.

We now return to the zero-point contribution to the grand
potential. As is well known for contact interactions, the
momentum sum in LHY correction has an ultraviolet divergence 
originating from the unphysical high-momentum behavior of the 
pseudopotential. This divergence is removed through the standard 
renormalization of the coupling constants,
\begin{equation}
g_{ij} \rightarrow g_{ij}
\Big(
1+
\frac{g_{ij}}{V}
\sum_{\mathbf{k}}
\frac{1}{2\epsilon_{\mathbf{k}}}
\Big),
\label{eq:renormalization}
\end{equation}
which expresses the bare interaction in terms of the physical
two-body scattering amplitude. Substituting the renormalized
couplings into Eq.~\eqref{eq:Bog-AB} and subsequently into
the LHY correction removes the ultraviolet divergence and yields 
the finite LHY contribution to the grand-potential density,
\begin{equation}
\label{eq:Bog-LHY}
\mathcal{E}_{\mathrm{LHY}}
=
\frac{1}{2}
\sum_{s\mathbf{k}}'
\Big(
E_{s\mathbf{k}}
-
A_{s\mathbf k}
+
\frac{g_s^2 n_0^2}
     {2\epsilon_{\mathbf{k}}}
\Big).
\end{equation}
The LHY term represents the leading beyond-mean-field
correction arising from quantum fluctuations and is essential
for the stabilization of quantum
droplets.

Having established the grand potential, we next determine the
equilibrium condensate density $n_0$ and chemical potential
$\mu$ from the coupled equations
\begin{equation}
\label{eq:self-cons}
\frac{\partial \mathcal{E}_{\mathrm{mf}}(n_0, \mu)}
{\partial n_0}\Big|_{\mu}
=
0,
\qquad
-\frac{1}{V}
\frac{\partial \mathcal{E}(n_0, \mu)}
{\partial \mu}\Big|_{n_0}
=
2n,
\end{equation}
where
$
\mathcal{E}
=
\mathcal{E}_{\mathrm{mf}}
+
\mathcal{E}_{\mathrm{LHY}}
$
is the renormalized grand potential. The first condition
determines the stationary condensate background, while the
second enforces particle-number conservation for the
two-component mixture. Together they determine $n_0$ and
$\mu$ self-consistently, including the effects of quantum
fluctuations contained in $\mathcal E_{\mathrm{LHY}}$. 
The condensate  condition is
equivalent to the Hugenholtz-Pines relation
\cite{hugenholtz_ground_1959}, which guarantees that the
density branch remains gapless by imposing
$
A_{+,\mathbf 0}-B_{+}=0
$
in Eq.~\eqref{eq:spectrumgen}. For the symmetric binary
mixture considered here, this yields
$
2gn_0+2g_{12}n_0-2\mu=0,
$
or equivalently
\begin{equation}
\label{eq:Bog-HP}
\mu= g_+ n_0.
\end{equation}
As a consequence, the density mode is gapless by
construction. Since
$
A_{-,\mathbf 0}=B_-=g_- n_0,
$
the spin branch is likewise gapless within standard
Bogoliubov theory. The mean-field relation between
$\mu$ and $n_0$ encodes the phononic character of the broken-symmetry
phase. 

Standard Bogoliubov theory is a
weak-coupling expansion rather than a self-consistent
scheme. Accordingly, fluctuation corrections
are evaluated by replacing $n_0\to n$ in the beyond-mean-field
expressions appearing in
Eqs.~\eqref{eq:Bog-AB}-\eqref{eq:Bog-LHY}. This procedure
maintains a consistent expansion in powers of $na^3$ and
avoids mixing contributions from different orders in the
dilute-gas expansion. For the same reason, the condensate
 conditon Eq.~\eqref{eq:self-cons} is
evaluated using only $\mathcal E_{\mathrm{mf}}$, while
$\mathcal E_{\mathrm{LHY}}$ enters through the number
equation and the resulting thermodynamic observables.
Using Eq.~\eqref{eq:Bog-HP} together with the replacement
$n_0\to n$ in fluctuation terms, the coefficient in
Eq.~\eqref{eq:Bog-AB} simplifies to
$
A_{s\mathbf k}
=
\epsilon_{\mathbf k}
+
g_s n,
$
yielding the  Bogoliubov spectrum
$
E_{s\mathbf k}
=
\sqrt{\epsilon_{\mathbf k}
(\epsilon_{\mathbf k}+2g_s n)},
$
for the density and spin branches. Substituting this result
into Eq.~\eqref{eq:Bog-LHY} gives the closed-form expression
for the LHY correction,
\begin{equation}
\frac{\mathcal{E}_{\mathrm{LHY}}}{V}
=
\frac{8}{15\pi^{2}}
\Big(\frac{m}{\hbar^{2}}\Big)^{3/2}
\big(
g_{+}^{5/2}
+
g_{-}^{5/2}
\big)
n^{5/2},
\end{equation}
which coincides with the symmetric-mixture limit of
Ref.~\cite{petrov_quantum_2015}.

We now turn to the second condition in
Eq.~\eqref{eq:self-cons}, which enforces particle-number
conservation.
This relation incorporates the effects of quadratic
fluctuations and therefore determines the condensate
depletion. Since $n_0$ and $\mu$ are treated as independent
variational parameters, the derivative with respect to
$\mu$ must be evaluated before imposing the
Hugenholtz-Pines condition in Eq.~\eqref{eq:Bog-HP}. This
procedure yields
\begin{equation}
\label{eq:generic-number-equation}
2n
=
2n_0
+
\frac{1}{2V}
\sum_{s\mathbf{k}}'
\Big(
\frac{A_{s\mathbf{k}}}{E_{s\mathbf{k}}}
-1
\Big).
\end{equation}
The second term gives the total quantum depletion of the
mixture and is equal to twice the depletion density of a
single component, denoted by $\tilde n$. Equation
\eqref{eq:generic-number-equation} is equivalent to the more familiar
derivation based on
$
\tilde n
=
V^{-1}
\sum_{\mathbf k}
\langle
c_{i\mathbf k}^{\dagger}
c_{i\mathbf k}
\rangle
-
n_0,
$
after expressing the particle operators in terms of
Bogoliubov quasiparticles and substituting the coefficients from Eq.~\eqref{eq:generic-usq-vsq}.
Resulting depletion is
\begin{equation}
\tilde n
=
\frac{1}{6\pi^{2}}
\left(\frac{m}{\hbar^{2}}\right)^{3/2}
\big(
g_{+}^{3/2}
+
g_{-}^{3/2}
\big)
n^{3/2}.
\end{equation}
Throughout this work, Bogoliubov thermodynamic quantities
are interpreted within an order-by-order weak-coupling
expansion. The replacement $n_0\to n$ reflects this
perturbative construction and should not be viewed as a
fully self-consistent determination of the condensate
density.

To facilitate a direct comparison with bosonic pairing theory,
which we introduce in the next section, it is useful to define
the interspecies anomalous average
\begin{equation}
\label{eq:generic-anom-av}
\tilde m_{12}
=
\frac{1}{V}
\sum_{\mathbf{k}}'
\langle
c_{1\mathbf{k}}
c_{2,-\mathbf{k}}
\rangle
=
-\frac{1}{2V}
\sum_{s\mathbf{k}}'
s
\frac{g_s n}
     {2E_{s\mathbf{k}}},
\end{equation}
where the condensate contribution is excluded from the summation. 
Physically, $\tilde m_{12}$ measures pairing correlations between 
the two species that originate from quantum fluctuations outside 
the condensate. Although standard Bogoliubov theory does not treat 
pairing as an independent variational degree of freedom, 
the anomalous average nevertheless quantifies the strength of 
interspecies pairing correlations. Similar to the LHY correction, 
the momentum sum appearing in Eq.~\eqref{eq:generic-anom-av} is 
ultraviolet divergent for a contact interaction. Restoring the 
renormalization term introduced in Eq.~\eqref{eq:renormalization} 
removes this divergence and leads to the finite result
\begin{equation}
\tilde m_{12}
=
\frac{1}{2\pi^{2}}
\Big(\frac{m}{\hbar^{2}}\Big)^{3/2}
\big(
g_{+}^{3/2}
-
g_{-}^{3/2}
\big)
n^{3/2}.
\end{equation}
It is further convenient to introduce the quantity
\begin{equation}
\Delta_{\mathrm{Bog}}
=
-\frac{g_{12}}{V}
\sum_{\mathbf{k}}
\langle
c_{1\mathbf{k}}
c_{2,-\mathbf{k}}
\rangle
=
-g_{12}
(
n_0+\tilde m_{12}
),
\end{equation}
which extends the anomalous average to include the condensate
contribution and scales it by the interspecies interaction.
For attractive interspecies interactions, $g_{12}<0$,
$\Delta_{\mathrm{Bog}}$ is positive by construction. Although
this quantity does not play the role of a self-consistent
variational parameter within Bogoliubov theory, it provides a natural
reference against which to compare the pairing field
$\Delta_0$ that emerges as the central parameter in
bosonic pairing theory.

Before turning to that theory, we briefly discuss the droplet
regime. For a homonuclear mixture with $g_{12}<-g$, the
mean-field contribution to the energy density becomes negative,
signaling an instability toward collapse. The beyond-mean-field
LHY correction introduces a repulsive contribution that can
stabilize the system and thereby enable the formation of
self-bound quantum droplets~\cite{petrov_quantum_2015}. At the same
time, however, the Bogoliubov spectrum develops an imaginary
density branch because $g_+=g+g_{12}$ becomes negative.
Consequently, quantities derived from the quasiparticle
spectrum acquire imaginary parts and can no longer be
interpreted straightforwardly within the quadratic theory.
Following Ref.~\cite{petrov_quantum_2015}, we therefore discard the
imaginary contribution associated with the unstable density
branch whenever $g_+<0$, by setting $g_+=0$. 
In the present work, this prescription is applied consistently to all 
fluctuation quantities that depend on the excitation spectrum, 
including the LHY correction, depletion density, anomalous average,
grand potential, and Bogoliubov coefficients
$u_{s\mathbf{k}}$ and $v_{s\mathbf{k}}$. This procedure
provides a practical extension of
Bogoliubov theory into the droplet regime and allows for a
meaningful comparison with bosonic pairing theory over all the 
parameter regime.

\section{Bosonic Pairing Theory}
\label{sec:pairing}

Bosonic pairing theory, introduced in
Refs.~\cite{hu_consistent_2020,hu_microscopic_2020}, provides an
alternative description of the binary Bose mixture in the
strongly attractive regime. Unlike standard Bogoliubov
theory, which requires the ad-hoc removal of the unstable
density branch in the droplet regime, bosonic pairing
yields a stable quasiparticle spectrum for all $g_{12}<0$.
At the thermodynamic level, however, both
theories attribute droplet stabilization to the competition
between attractive mean-field interactions and repulsive
LHY corrections. Below, we reproduce  the bosonic pairing theory in the 
operator formalism for a homogeneous binary Bose mixture.

Following Ref.~\cite{hu_consistent_2020}, we start with the exact Hamiltonian 
from Eq.~\eqref{eq:generic-exactH}, with equal masses, equal intraspecies 
interactions, and equal total particle densities. We first consider the 
intraspecies terms
\begin{align} \label{eq:generic-intraspecies-Hamiltonian}
    \mathcal{H}_{ii} =  \sum_{\mathbf{k}}
(\epsilon_{\mathbf{k}}-\mu)
c_{i\mathbf{k}}^{\dagger}c_{i\mathbf{k}}
 + \frac{g}{2V}
\sum_{\mathbf{k}\mathbf{k}'\mathbf{q}}
c_{i,\mathbf{k}+\mathbf{q}}^{\dagger}
c_{i,\mathbf{k}'-\mathbf{q}}^{\dagger}
c_{i \mathbf{k}'} c_{i \mathbf{k}}
\end{align}
which include the kinetic single-particle term and the intraspecies interaction 
term, and apply the condensate approximation $c_{i\mathbf 0}\approx\sqrt{N_0}$. 
Substituting back into the above equation and expanding up to quadratic order 
in fluctuations $c_{i, \mathbf k \neq \mathbf 0}$, we get the condensate and 
quadratic contributions to the Hamiltonian
\begin{equation} \label{eq:BP-intraspecies-Hamiltonian}
\begin{aligned}
    \mathcal{H}_{ii} &\simeq V (gn_0^2 -2\mu n_0) \\& 
    +\sum_\mathbf k ' \Big[ \bar{\xi}_\mathbf k c_{i\mathbf k}^{\dagger}c_{i\mathbf k}
+\frac{gn_0}{2}
\big(c_{i\mathbf k}^{\dagger}c_{i,-\mathbf k}^{\dagger}
+\mathrm{H.c.}\big) \Big]
\end{aligned}
\end{equation}
where
$
\bar{\xi}_\mathbf k = \epsilon_{\mathbf k} - \mu + 2gn_0.
$
The first term is the condensate contribution to the grand potential from 
intraspecies terms. The treatment of the intraspecies terms in the exact 
Hamiltonian is the same for both Bogoliubov theory and bosonic pairing theory.

To identify the approximation underlying bosonic pairing,
we rewrite the interspecies interaction term
\begin{align}
  \mathcal H_{12}  =
\frac{g_{12}}{V} \sum_{\mathbf{k}\mathbf{k}'\mathbf{q}}
c_{1,\mathbf{k}+\mathbf{q}}^{\dagger}
c_{2,\mathbf{k}'-\mathbf{q}}^{\dagger}
c_{2 \mathbf{k}'} c_{1 \mathbf{k}}
\end{align}
in terms of the pair annihilation operator
$
P_{\mathbf q}=\sum_{\mathbf k}
c_{1,\mathbf q+\mathbf k}c_{2,-\mathbf k},
$
where $\mathbf q$ denotes the center-of-mass momentum of the pair. 
The interaction Hamiltonian can then be expressed as
$
\mathcal H_{12}
=
\frac{g_{12}}{V}
\sum_{\mathbf q}
P_{\mathbf q}^{\dagger}P_{\mathbf q}.
$
We then introduce the pairing mean field
\begin{equation}
\label{eq:BP-pair-field-general}
\Delta_{\mathbf q}
\equiv
-\frac{g_{12}}{V}\langle P_{\mathbf q}\rangle
=
-\frac{g_{12}}{V}
\sum_{\mathbf k}
\langle
c_{1,\mathbf q+\mathbf k}
c_{2,-\mathbf k}
\rangle,
\end{equation}
which includes all momenta, including the condensate mode.
We decompose the pair operator as
$
P_{\mathbf q}
=
-\frac{V}{g_{12}}\Delta_{\mathbf q}
+
\delta P_{\mathbf q},
$
where $\delta P_{\mathbf q}$ denotes fluctuations around the mean.
Substituting this decomposition into $\mathcal H_{12}$ yields the exact rewriting
\begin{align}
\mathcal H_{12}
=&
-
V\sum_{\mathbf q}
\frac{|\Delta_{\mathbf q}|^2}{g_{12}}
-
\sum_{\mathbf q}
\left(
\Delta_{\mathbf q}P_{\mathbf q}^{\dagger}
+
\Delta_{\mathbf q}^{*}P_{\mathbf q}
\right)
\nonumber\\
&
+
\frac{g_{12}}{V}
\sum_{\mathbf q}
\delta P_{\mathbf q}^{\dagger}
\delta P_{\mathbf q}.
\label{pair_exact_decomposition}
\end{align}
The bosonic-pairing approximation consists of retaining only the quadratic
contribution of the attractive interspecies interaction while neglecting the
higher-order fluctuation term proportional to
$\delta P_{\mathbf q}^{\dagger}\delta P_{\mathbf q}$.
We set
$
\Delta_{\mathbf q}
=
\Delta_0 \delta_{\mathbf q\mathbf 0},
$
which is equivalent to setting the pairing field to a uniform saddle-point 
solution in the position-space representation in Ref.~\cite{hu_microscopic_2020}.
Taking $\Delta_0$ to be real without loss of generality and using
Eq.~\eqref{eq:generic-anom-av}, we obtain
$
\Delta_0
=
-g_{12}
(n_0+\tilde m_{12}),
$
where the condensate contribution arises because the pair field includes the
$\mathbf k=\mathbf 0$ mode.
Substituting this expression back into the interspecies Hamiltonian
gives the reduced form
$
\mathcal H_{12}
\simeq
-
V \Delta_0^2 / g_{12}
-
\Delta_0 P_{\mathbf 0}^{\dagger} + \Delta_0 P_{\mathbf 0},
$
where all terms with $\mathbf q \neq \mathbf 0$ are treated as higher-order
fluctuations and are neglected at the pairing mean-field level.

This decomposition allows us to express the pair operator in terms of 
$c_{i,\mathbf{k}\neq 0}$ and diagonalize the Hamiltonian, yielding
\begin{equation} \label{eq:BP-Hamiltonian-final}
\frac{\mathcal H_{12}}{V}
\simeq
-\frac{\Delta_0^2}{g_{12}}
-2\Delta_0 n_0
-
\frac{\Delta_0}{V} \sum_{\mathbf k}'
\big(
c_{1\mathbf k}^{\dagger}
c_{2,-\mathbf k}^{\dagger}
+\mathrm{H.c.}
\big),
\end{equation}
for the symmetric mixture. The second term originates from the condensate-mode
contribution extracted from the $P_{\mathbf 0}$ terms. The resulting quadratic
Hamiltonian can be cast into BdG form, as in Eq.~\eqref{eq:generic-BdG-form},
with the matrix
\begin{equation}
\mathbb M_{\mathbf k}
=
\begin{pmatrix}
\bar{\xi}_\mathbf k & 0 & gn_0 & -\Delta_0 \\
0 & \bar{\xi}_\mathbf k & -\Delta_0 & gn_0 \\
gn_0 & -\Delta_0 & \bar{\xi}_{-\mathbf k} & 0 \\
-\Delta_0 & gn_0 & 0 & \bar{\xi}_{-\mathbf k}
\end{pmatrix}.
\label{eq:BP-BdG-matrix}
\end{equation}
The corresponding mean-field grand potential is
\begin{equation} \label{eq:BP-MF-grand-potential}
\frac{\mathcal E_{\mathrm{mf}}}{V}
=
gn_0^2
-
2\mu n_0
-
\frac{\Delta_0^2}{g_{12}}
-
2\Delta_0 n_0.
\end{equation}
The first two terms originating from Eq.~\eqref{eq:BP-intraspecies-Hamiltonian} 
coincide with those of standard Bogoliubov theory,
as can be seen by comparison with Eq.~\eqref{eq:Bog-MF-grand-potential},
whereas the last two terms arise from the interspecies pairing channel 
as Eq.~\eqref{eq:BP-Hamiltonian-final}. In the limit where $\tilde{m}_{12}$ 
is neglected in its contribution to the mean-field
grand potential, the interspecies mean-field terms reduce to the same 
structure as Bogoliubov theory's. This can be seen by substituting
$\Delta_0 = -g_{12} n_0$ into Eq.~\eqref{eq:BP-MF-grand-potential}.

Proceeding as in the previous section and diagonalizing the quadratic
Hamiltonian, we obtain the quasiparticle spectrum
\begin{equation} \label{eq:BP-spectrum}
E_{s\mathbf k}
=
\sqrt{
\bar{\xi}_{\mathbf k}^{2}
-
(gn_0 - s\Delta_0)^2
},
\end{equation}
which has the generic BdG form of Eq.~\eqref{eq:BdG-generic}. 
The corresponding coefficients are
\begin{equation} \label{eq:BP-AB}
A_{s\mathbf k} = \bar{\xi}_{\mathbf k},
\quad
B_s = gn_0 - s\Delta_0,
\end{equation}
with $A_{s\mathbf k}$ now independent of the branch index $s$.
We next derive the thermodynamic quantities within bosonic pairing
theory. After ultraviolet renormalization, the LHY-like contribution
to the grand potential is
\begin{equation} \label{eq:BP-LHY}
\mathcal E_{\mathrm{LHY}}
=
\frac{1}{2}
\sum_{s\mathbf k}'
\Big[
E_{s\mathbf k}
-
\bar{\xi}_\mathbf k
+
\frac{(gn_0-s\Delta_0)^2}{2\epsilon_{\mathbf k}}
\Big].
\end{equation}
Having obtained the mean-field grand potential, the quasiparticle
spectrum, and the LHY contribution, we determine the chemical
potential $\mu$, the condensate density $n_0$, and the pairing field
$\Delta_0$ self-consistently.

We start from the condensate-density condition
$
\frac{\partial \mathcal{E}_{\mathrm{mf}}}{\partial n_0} \big|_{\Delta_0, \mu} = 0,
$
as in Eq.~\eqref{eq:self-cons}, which determines the chemical potential 
in terms of the condensate density and pairing field,
\begin{equation} \label{eq:BP-chemical-potential}
\mu = g n_0 - \Delta_0 = g_+ n_0 + g_{12}\tilde{m}_{12}.
\end{equation}
Using this relation, the coefficient $A_{s\mathbf k}$ and with 
it $\bar{\xi}_\mathbf k$ take the simplified form
$
A_{s\mathbf k}
=
\epsilon_{\mathbf k}
+
g n_0
+
\Delta_0.
$
The quasiparticle spectrum, in agreement with Ref.~\cite{hu_microscopic_2020}, 
then separates into a density branch ($s=+$),
$
E_{+,\mathbf k}
=
\sqrt{
(\epsilon_{\mathbf k}+2g n_0)
(\epsilon_{\mathbf k}+2\Delta_0)
},
$
and a spin branch ($s=-$) 
$
E_{-,\mathbf k}
=
\sqrt{
\epsilon_{\mathbf k}
(\epsilon_{\mathbf k}+2g n_0+2\Delta_0)
}.
$
The density mode is therefore gapped, whereas the spin mode remains gapless. 
This should be contrasted with standard Bogoliubov theory, where both collective 
modes are gapless. Notably, the density branch that becomes gapped here is the 
same mode that turns dynamically unstable and develops an imaginary dispersion 
in the droplet regime within Bogoliubov theory.

We now turn to the self-consistent determination of the particle density. 
Equation~\eqref{eq:generic-number-equation} provides a general expression 
valid in both descriptions, since the chemical potential always enters 
as an intraspecies normal contribution. Substituting Eq.~\eqref{eq:BP-AB} 
yields bosonic pairing's number equation which, 
together with Eq.~\eqref{eq:BP-chemical-potential}, forms 
the first two of the set of self-consistency relations for $n_0$ and $\Delta_0$:
\begin{equation}
\label{eq:BP-number-equation}
n
=
n_0
+
\frac{1}{2V}
\sum_{s\mathbf{k}}'
\frac{1}{2}\Big(
\frac{\epsilon_{\mathbf k} + gn_0 + \Delta_0}{E_{s\mathbf{k}}}
-1
\Big),
\end{equation}
in which the numerator of the first term is branch-independent. 
The pairing field $\Delta_0$ introduces an additional
stationarity condition for the grand potential
$\mathcal E(n_0,\mu,\Delta_0,V)$ beyond those in
Eq.~\eqref{eq:self-cons}, namely 
$
\frac{\partial \mathcal E}
{\partial \Delta_0}
\big|_{\mu,n_0}
=
0,
$
which yields the equation for the pairing field
\begin{equation}
\label{eq:BP-anom-av}
\Delta_0 + g_{12} n_0
=
\frac{g_{12}}{4V}
\sum_{s \mathbf{k}}'
\left(sgn_0-\Delta_0\right)
\Big(
\frac{1}{E_{s\mathbf{k}}}
-
\frac{1}{\epsilon_{\mathbf{k}}}
\Big).
\end{equation}
The ultraviolet divergence in Eq.~\eqref{eq:BP-anom-av} has already 
been removed in $\mathcal E_{\mathrm{LHY}}$, Eq.~\eqref{eq:BP-LHY}. 
Varying the grand potential 
with respect to the pairing field while fixing the chemical potential 
and the condensate density is equivalent to varying it with respect 
to the anomalous average $\tilde{m}_{12}$, as 
the pairing field can be decomposed into a condensate contribution 
and a fluctuation contribution
$
\Delta_0=-g_{12}(n_0+\tilde m_{12}).
$

Essentially, bosonic pairing theory elevates
Eq.~\eqref{eq:generic-anom-av} to a self-consistency equation, so
that the anomalous average co-determines thermodynamic quantities
such as the grand potential and the quasiparticle spectrum. It also
modifies the condensate density $n_0$, since
Eqs.~\eqref{eq:generic-number-equation} and
\eqref{eq:BP-anom-av} become coupled self-consistency equations.
Furthermore, the spectrum in Eq.~\eqref{eq:BP-spectrum}, which
enters the higher-order LHY correction to the condensate density, is
itself modified.
In this respect, bosonic pairing theory is reminiscent of
Hartree-Fock-Bogoliubov (HFB) theory
\cite{griffin_conserving_1996,giorgini_thermodynamics_1997,hutchinson_finite-temperature_1997,hu_consistent_2020},
where both the normal and anomalous averages are included in the
quadratic Hamiltonian and determined self-consistently. Bosonic
pairing theory differs from HFB, however, in that it retains only the
interspecies anomalous channel.  Interestingly, the widely used Popov
approximation to HFB makes the opposite choice, retaining the normal
contribution while neglecting the anomalous density.

The bosonic pairing approximation at the Hamiltonian level warrants 
closer examination. The quadratic Hamiltonian in 
Eq.~\eqref{eq:BP-Hamiltonian-final} has a BCS-like structure in which 
the interspecies attraction is projected entirely onto the 
zero-center-of-mass pairing channel. Consequently, interspecies anomalous 
pairing processes are retained through $\Delta_0$. At the same time, 
the bosonic pairing BdG matrix in Eq.~\eqref{eq:BP-BdG-matrix} shows 
that normal interspecies terms of the form
$
c_{1\mathbf k}^{\dagger}c_{2\mathbf k}
$
are entirely absent, unlike in the Bogoliubov matrix of 
Eq.~\eqref{eq:Bog-BdG-matrix}. To understand the origin of these 
missing contributions, consider the finite-momentum pair-field
fluctuations,
$
\delta P_{\mathbf q\neq\mathbf 0}
=
P_{\mathbf q\neq\mathbf 0}.
$
To leading order in the condensate amplitude,
$
\delta P_{\mathbf q\neq\mathbf 0}
\simeq
\sqrt{N_0}
(c_{1\mathbf q}+c_{2\mathbf q}),
$
showing that finite-momentum pairs are directly coupled to
single-particle excitations through the condensate.
Consequently, the fluctuation term omitted in the pairing
mean-field approximation generates, at quadratic order,
\begin{equation} \label{eq:BP-fluctuation-corrections}
\begin{aligned}
    \frac{g_{12}}{V}
\sum_{\mathbf q}'
\delta P_{\mathbf q}^{\dagger}
\delta P_{\mathbf q}
\simeq
g_{12}n_0
\sum_{\mathbf q}'
\Big(
c_{1\mathbf q}^{\dagger}c_{2\mathbf q}
+
c_{2\mathbf q}^{\dagger}c_{1\mathbf q}
+ \sum_i c_{i\mathbf q}^{\dagger}c_{i\mathbf q}
\Big).
\end{aligned}
\end{equation}
These are precisely the normal interspecies contributions
retained in Bogoliubov theory but omitted in the
bosonic-pairing approximation.

The connection to Bogoliubov theory becomes more transparent
in the density-spin basis. In this representation, the
neglected finite-momentum pair fluctuations contribute
$
\frac{g_{12}}{V}\sum_{\mathbf q}'
\delta P_{\mathbf q}^{\dagger}\delta P_{\mathbf q}
\simeq
\sum_{s\mathbf k}'
\delta A_{s\mathbf k}
 c_{s\mathbf k}^{\dagger}c_{s\mathbf k},
$
with
$
\delta A_{s\mathbf k}=(1+s)g_{12}n_0.
$
The omitted term therefore acts entirely in the normal
channel and renormalizes the density ($s=+1$) branch,
while leaving the spin ($s=-1$) branch unchanged.
Importantly, this correction is of the same quadratic
order as the terms retained in the bosonic-pairing
Hamiltonian. Restoring it modifies the coefficients to
$
A_{s\mathbf k}
=
\bar{\xi}_\mathbf k
+
(1+s)g_{12}n_0,
$
and
$
B_s
=
g_s n_0
+
s g_{12}\tilde m_{12}.
$
These coefficients have the same structure as those
obtained in Bogoliubov theory as Eq.~\eqref{eq:Bog-AB} shows.

Consequently, once finite-momentum pair fluctuations are included, the only
remaining difference between the two approaches is the anomalous-average
contribution $\tilde{m}_{12}$ to $\Delta_0$, and consequently to the
mean-field grand potential in
Eq.~\eqref{eq:BP-MF-grand-potential} and the quasiparticle spectrum in
Eq.~\eqref{eq:BP-spectrum}.
The essential approximations underlying the bosonic
pairing theory are the restriction of the attractive interspecies interaction
to the zero-center-of-mass pairing channel together with the neglect of
finite-momentum pair fluctuations. In the remainder of this paper, however,
we neglect the pairing-fluctuation corrections in
Eq.~\eqref{eq:BP-fluctuation-corrections} and instead follow the original
formulation of Ref.~\cite{hu_consistent_2020} when solving the bosonic-pairing
equations. Nevertheless, the shift in the coefficient
$A_{s\mathbf{k}}$  eliminates the imaginary branch of the excitation spectrum.
Moreover, derived quantities such as the LHY energy correction
$\mathcal{E}_{\mathrm{LHY}}$ and the per-species condensate density $n_0$
remain finite after droplet formation, $g_{12}+g\le0$.

In the bosonic pairing theory, the density branch of the spectrum 
is gapped. Setting $\mathbf{k}=\mathbf{0}$ yields
$
E_{+,\mathbf{0}} = 2\sqrt{gn_0\Delta_0},
$
which, unlike the BCS result~\cite{fetter_quantum_1971}, is not directly
proportional to the magnitude of the pairing field. Furthermore, this
spectral gap is not generated solely by the anomalous-average contribution
$\tilde{m}_{12}$, as in HFB theory~\cite{griffin_conserving_1996}. Instead,
it originates primarily from restricting the attractive interaction to the
zero-center-of-mass pairing channel. This becomes evident by setting
$\tilde{m}_{12}=0$, which gives
$
E_{+,\mathbf{0}}
\approx
2n_0\sqrt{g|g_{12}|}.
$

Following~\cite{hu_consistent_2020}, 
we solve for the condensate density self-consistently. In other 
words, we do not make the substitution $n_0 \rightarrow n$ for 
beyond-mean-field expressions. However, we note that if one formulates 
the theory in terms of an order-by-order expansion in $na^3$, and 
therefore also ignores the contributions of the interspecies anomalous 
average $\tilde{m}_{12}$ , 
one could further simplify many of the above expressions to obtain for example
$
\Delta_0\approx -g_{12}n.
$
Moreover, Eqs.~\eqref{eq:generic-number-equation} and \eqref{eq:BP-anom-av} 
for the depletion density and the interspecies anomalous average 
would then reduce to analytic integrals in the continuum limit. 
In this treatment, the spin branch becomes identical in form to the Bogoliubov 
result, while the density branch is still gapped. Such a non-self consistent 
calculation yields values which can serve as initial guesses for the 
self-consistency equations.

\begin{figure}[htbp]
    \centering
    \includegraphics[width=\columnwidth]{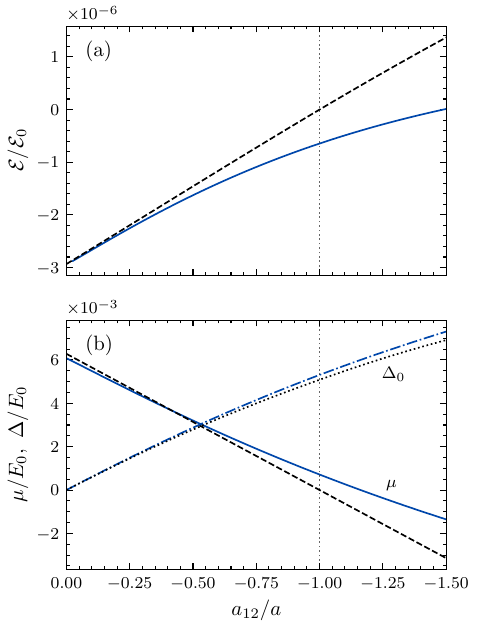}
\caption{(a) Grand-potential density as a function of the interaction-strength 
ratio $a_{12}/a$, spanning the attractive mean-field-stable regime and 
extending into the droplet regime. 
The blue solid curve denotes bosonic pairing theory, and the black dashed 
curve denotes Bogoliubov theory. 
Values in panel (a) are measured in units of $\mathcal{E}_0 = \hbar^2/(ma^5)$. 
(b) Chemical potential $\mu$ and pairing field $\Delta_0$ as 
functions of $a_{12}/a$ across the mean-field-to-droplet crossover. 
The blue solid and blue dash-dotted curves denote $\mu$ and $\Delta_0$ 
in bosonic pairing theory, respectively, while the black dashed and 
black dotted curves denote the corresponding Bogoliubov results. 
Values in panel (b) are measured in units of $E_0 = \hbar^2/(ma^2)$. 
In both panels, the per-species density is fixed at $na^3 = 0.5\times10^{-3}$. }
    \label{fig:combined}
\end{figure}
\begin{figure}[htbp]
    \centering
    \includegraphics[width=\columnwidth]{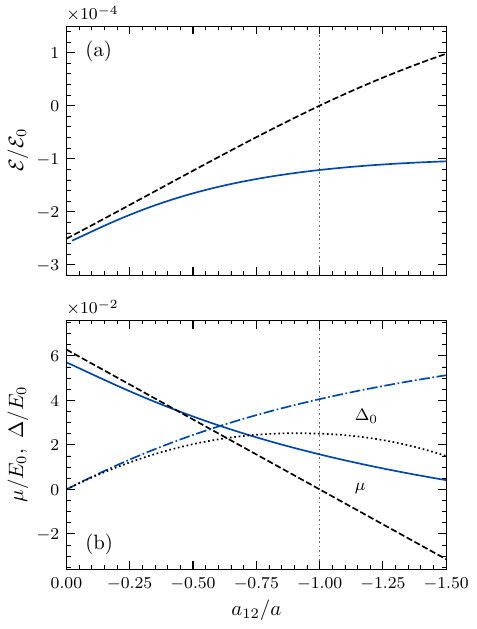}
\caption{Same as Fig.~\ref{fig:combined}, but for a tenfold higher density. 
In both panels, the per-species density is fixed at $na^3 = 0.5\times10^{-2}$. 
The differences between the Bogoliubov and bosonic pairing theories become 
more pronounced at higher density, particularly in the droplet regime.}
\label{fig:combined-high-density}
\end{figure}

In the bosonic pairing theory, the quasiparticle spectrum remains real for 
all attractive interactions $a_{12}<0$. Consequently, the point 
$a_{12}/a=-1$ is non-singular, allowing us to analyze the theory throughout 
the mean-field stable crossover regime. A key difference 
in determining the variational parameters is that, in bosonic pairing theory, 
the number equation and the gap equation are coupled and must be solved 
self-consistently. In this work, we solve Eqs.~\eqref{eq:generic-number-equation} 
and~\eqref{eq:BP-anom-av} numerically.
We emphasize that, while previous studies of bosonic pairing have relaxed 
the number constraint, we enforce fixed particle number as expressed 
in Eq.~\eqref{eq:generic-number-equation}~\cite{hu_consistent_2020,hu_microscopic_2020}. 
This choice is motivated by the scope of our analysis, which is not 
restricted to the droplet regime but instead addresses the crossover from 
the gas to the droplet regime, thereby more closely following the experimental 
pathway to droplet formation.

\begin{figure}[htbp]
    \centering
    \includegraphics[width=\columnwidth]{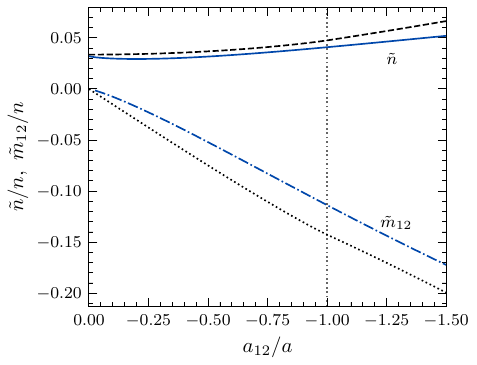}
    \caption{ Intraspecies normal fraction $\tilde n/n$ and interspecies 
    anomalous fraction $\tilde m_{12}/n$ as functions of the 
    interaction-strength ratio $a_{12}/a$, spanning the attractive 
    mean-field-stable regime and extending into the droplet regime.
    The blue solid and black dashed curves denote $\tilde n/n$ in 
    bosonic pairing theory and Bogoliubov theory, respectively. 
    The blue dash-dotted and black dotted curves denote the corresponding 
    results for $\tilde m_{12}/n$. 
    The per-species density is fixed at $na^3 = 0.5\times10^{-3}$. }
\label{fig:density-comparison}
\end{figure}

Figure~\ref{fig:combined} shows that the variationally minimized 
ground-state grand potential, 
$
\mathcal{E} = \mathcal{E}_{\mathrm{mf}} + \mathcal{E}_{\mathrm{LHY}},
$
is lower in bosonic pairing theory than in Bogoliubov theory. 
As expected, both approaches reproduce identical single-species results 
in the limit $a_{12}\rightarrow 0$ for all thermodynamic expectation 
values. For repulsive interspecies interactions, the 
$\mathbf{k}=\mathbf{0}$ quasiparticle gap in bosonic pairing theory becomes 
imaginary when $\Delta_0<0$, signaling a dynamical instability. We therefore 
exclude the region $a_{12}>0$, where the bosonic pairing framework 
is not applicable to mixtures with repulsive interspecies interactions. 

Figure~\ref{fig:combined}(b) highlights another important feature: 
the Bogoliubov quantity $\Delta_{\mathrm{Bog}}$ and the bosonic pairing 
field $\Delta_0$ exhibit an average relative deviation of approximately 
$0.030$ in the mean-field stable regime and approximately $0.053$ in 
the droplet regime, while converging to the same value as $a_{12}\to 0$. 
The same limiting agreement is observed for 
$\mu$. A key difference is that, within Bogoliubov theory, the chemical 
potential crosses zero exactly at the mean-field instability point, whereas 
in bosonic pairing theory one has $\mu = g_{12}\tilde{m}_{12}$ at the 
onset of droplet formation, which can also be seen from 
Eq.~\eqref{eq:BP-chemical-potential}.

Since the droplet density is higher than that of the gaseous phase and
depends on the experimental setup
\cite{cabrera_quantum_2018,semeghini_self-bound_2018},
we solve the self-consistency equations at density
$na^3 = 0.5 \times 10^{-2}$, which is an order of magnitude larger than
the density considered in Fig.~\ref{fig:combined}.
Figure~\ref{fig:combined-high-density}(a) shows the variationally
minimized grand potential, where bosonic pairing theory again yields a
lower value than Bogoliubov theory at the higher density. The relative
difference between the two theories is also enhanced at higher density.
As in Fig.~\ref{fig:combined}, this difference increases as the
interspecies scattering length decreases.
Figure~\ref{fig:combined-high-density}(b) shows the chemical potential
together with $\Delta_{\mathrm{Bog}}$ and $\Delta_0$, whose difference
likewise increases with decreasing $a_{12}$.

Figure~\ref{fig:density-comparison} shows that the discrepancy between the
two theories is already visible in the per-species depletion density in the
mean-field-stable regime, where the relative difference between 
each theory's average depletion densities over the regime is
$\delta\tilde{n}/\tilde{n}\approx 0.11$. This discrepancy becomes larger in
the droplet regime, increasing to approximately $0.25$.
In Petrov’s single-branch prescription, 
the contribution from the unstable density branch 
$
\sqrt{\epsilon_{\mathbf{k}}( \epsilon_{\mathbf{k}} + 2g_+ n )}
$ 
is omitted in the evaluation of thermodynamic quantities. 
In contrast, bosonic pairing theory retains the gapped branch, which 
suppresses the depletion density in the droplet regime.
The anomalous average $\tilde{m}_{12}$ is negative for the 
interspecies-attractive mixture and increases in magnitude as the 
attraction is strengthened in both theories. In bosonic pairing theory, 
the deviation of $\tilde{m}_{12}$ from the Bogoliubov result is more 
pronounced in both the mean-field stable and droplet regimes. 
We attribute this difference to the self-consistent inclusion of the 
anomalous average in bosonic pairing theory. 
For $\tilde{m}_{12}$, however, the change in the slope at the onset 
of droplet formation partially reduces the difference in magnitude 
between the two theories.

\section{Two-Body Correlations}
\label{sec:corr}

The thermodynamic comparison of the previous section shows that
bosonic pairing theory is variationally favored, but does not
directly reveal how the gapped density branch manifests in
observables. Two-body correlation functions provide this link,
since they retain momentum-resolved information about the
collective excitation spectrum. We focus on interspecies
correlations, as these are most directly affected by the
differing treatments of the interspecies attraction.
Within both theories, the correlation functions follow exactly
from the Bogoliubov coefficients $u_{s\mathbf{k}}$ and
$v_{s\mathbf{k}}$, and therefore depend on the two approaches
only through the coefficients $A_{s\mathbf{k}}$ and $B_s$.

\subsection{Momentum-space correlations}

Momentum-space two-body correlations provide a direct probe of
pairing and collective fluctuations in binary Bose mixtures.
Within the Gaussian theories considered here, they can be
evaluated exactly and expressed entirely in terms of normal
and anomalous averages. We begin from the momentum-resolved
second-order correlation function~\cite{glauber_quantum_1963}
\begin{align}
G_{ij}^{(2)}(\mathbf{k},\mathbf{k}')
=
\langle
c_{i\mathbf{k}}^{\dagger}
c_{j\mathbf{k}'}^{\dagger}
c_{i\mathbf{k}}
c_{j\mathbf{k}'}
\rangle.
\end{align}
Using the Bogoliubov transformation introduced in
Sec.~\ref{sec:bog},
$
c_{s\mathbf{k}}
=
u_{s\mathbf{k}}b_{s\mathbf{k}}
+
v_{s\mathbf{k}}b_{s,-\mathbf{k}}^{\dagger},
$
together with the quasiparticle vacuum condition
$
b_{s\mathbf{k}}|0\rangle=0,
$
all higher-order expectation values can be reduced to
products of two-point functions. Equivalently, Wick's
theorem holds exactly for the quadratic theories considered
in this work. For the interspecies correlator we obtain
$
G_{12}^{(2)}(\mathbf{k},\mathbf{k}')
=
\langle
c_{1\mathbf{k}}^{\dagger}
c_{2\mathbf{k}'}^{\dagger}
\rangle
\langle
c_{1\mathbf{k}}
c_{2\mathbf{k}'}
\rangle
+
\langle
c_{1\mathbf{k}}^{\dagger}
c_{1\mathbf{k}}
\rangle
\langle
c_{2\mathbf{k}'}^{\dagger}
c_{2\mathbf{k}'}
\rangle
+
\langle
c_{1\mathbf{k}}^{\dagger}
c_{2\mathbf{k}'}
\rangle
\langle
c_{2\mathbf{k}'}^{\dagger}
c_{1\mathbf{k}}
\rangle .
$
For a symmetric mixture,
$
n_{1\mathbf{k}}
=
n_{2\mathbf{k}}
\equiv
n_{\mathbf{k}},
$
so that
\begin{equation}
G_{12}^{(2)}(\mathbf{k},\mathbf{k}')
=
|\langle
c_{1\mathbf{k}}^{\dagger}
c_{2\mathbf{k}'}^{\dagger}
\rangle|^2
+
n_{\mathbf{k}}n_{\mathbf{k}'}
+
|\langle
c_{1\mathbf{k}}^{\dagger}
c_{2\mathbf{k}'}
\rangle|^2.
\end{equation}
The three terms have a simple interpretation: they originate
from interspecies anomalous pairing correlations,
uncorrelated density fluctuations, and interspecies normal
coherences, respectively.
For a homogeneous system, translational invariance imposes
momentum conservation,
$
\langle
c_{i\mathbf{k}}^{\dagger}
c_{j\mathbf{k}'}
\rangle
=
\delta_{\mathbf{k} \mathbf{k}'}
\langle
c_{i\mathbf{k}}^{\dagger}
c_{j\mathbf{k}}
\rangle,
$
and
$
\langle
c_{i\mathbf{k}}
c_{j\mathbf{k}'}
\rangle
=
\delta_{\mathbf{k},-\mathbf{k}'}
\langle
c_{i\mathbf{k}}
c_{j,-\mathbf{k}}
\rangle.
$
Consequently, correlations at equal momenta
$(\mathbf{k},\mathbf{k})$ probe normal fluctuations,
whereas correlations between opposite momenta
$(\mathbf{k},-\mathbf{k})$ probe pairing correlations.

To simplify the remaining discussion, we first introduce the 
shorthand notations for
\begin{align}
n_{\mathbf{k}}
&=
\langle c^\dagger_{i\mathbf{k}}c_{i\mathbf{k}}\rangle
=
\frac{1}{4}
\sum_{s}
\Big(
\frac{A_{s\mathbf{k}}}{E_{s\mathbf{k}}}
-1
\Big), 
\label{eq:generic-correlation-coefficient-1}
\\
n_{12,\mathbf{k}}
&=
\langle c^\dagger_{1\mathbf{k}}c_{2\mathbf{k}}\rangle
=
\frac{1}{4}
\sum_{s}
s
\frac{A_{s\mathbf{k}}}{E_{s\mathbf{k}}},
\\
m_{\mathbf{k}}
&=
\langle c_{i\mathbf{k}}c_{i,-\mathbf{k}}\rangle
=
-\frac{1}{4}
\sum_{s}
\frac{B_s}{E_{s\mathbf{k}}},
\\
m_{12,\mathbf{k}}
&=
\langle c_{1\mathbf{k}}c_{2,-\mathbf{k}}\rangle
=
-\frac{1}{4}
\sum_{s}
s
\frac{B_s}{E_{s\mathbf{k}}},
\label{eq:generic-correlation-coefficient-4}
\end{align}
where $s= \pm 1$ labels the density and spin branches, respectively. 
We then follow Refs.~\cite{tenart_observation_2021,butera_position-_2021}, 
and separate the interspecies correlations into normal and anomalous channels,
\begin{gather}
\label{eq:norm-corr-k}
G^{(2)}_{12,\mathrm{A}}(\mathbf{k}) = G_{12}^{(2)}(\mathbf k, -\mathbf k) 
=
n_{\mathbf{k}}^2
+
m_{12,\mathbf{k}}^2,
\\
G^{(2)}_{12,\mathrm{N}}(\mathbf{k}) = G_{12}^{(2)}(\mathbf k, \mathbf k) 
=
n_{\mathbf{k}}^2
+
n_{12,\mathbf{k}}^2.
\label{eq:anom-corr-k}
\end{gather}
The anomalous contribution isolates correlations between
opposite-momentum particles and therefore directly probes
pair formation, whereas the normal contribution measures
equal-momentum interspecies coherence. 
Similarly, for intraspecies correlations,
$
G^{(2)}_{ii,\mathrm{A}}(\mathbf{k})
=
n_{\mathbf{k}}^2+m_{\mathbf{k}}^2
$
and
$
G^{(2)}_{ii,\mathrm{N}}(\mathbf{k})
=
2n_{\mathbf{k}}^2.
$
Since the mixture is symmetric, the latter two correlations are 
identical for both species.

For comparison with experiment, it is convenient to
introduce the normalized correlation functions~\cite{naraschewski_spatial_1999}
\begin{equation}
h^{(2)}_{ij}(\mathbf{k},\mathbf{k}')
=
\frac{
G^{(2)}_{ij}(\mathbf{k},\mathbf{k}')
}
{
n_{\mathbf{k}}n_{\mathbf{k}'}
}
-1.
\end{equation}
The equal-momentum and opposite-momentum limits,
$
h^{(2)}_{ij,\mathrm{N}}(\mathbf{k})
=
h^{(2)}_{ij}(\mathbf{k},\mathbf{k})
$
and
$
h^{(2)}_{ij,\mathrm{A}}(\mathbf{k})
=
h^{(2)}_{ij}(\mathbf{k},-\mathbf{k}),
$
therefore provide direct measures of normal and pairing
correlations, respectively.
The numerical values of these correlators are controlled by the
amplitudes $A_{s\mathbf k}$, $B_s$, and the quasiparticle spectrum
$E_{s\mathbf k}$, which are themselves determined by the
self-consistency equations: Eq.~\eqref{eq:self-cons} for
Bogoliubov theory, and Eqs.~\eqref{eq:BP-number-equation} and
\eqref{eq:BP-anom-av} for bosonic pairing theory.

\subsection{Position-space correlations}

While momentum-space correlations probe pairing between specific
momentum states, position-space correlations characterize the
spatial structure of density and pairing fluctuations. We define
the normalized second-order correlation function as~\cite{glauber_quantum_1963}
\begin{equation}
\mathcal{G}_{ij}^{(2)}(\mathbf r,\mathbf r')
=
\frac{1}{n^2}
\big\langle
\psi_i^\dagger(\mathbf r)
\psi_j^\dagger(\mathbf r')
\psi_i(\mathbf r)
\psi_j(\mathbf r')
\big\rangle,
\end{equation}
where $\psi_i(\mathbf r)$ is the bosonic field operator in
position space. We employ this normal-ordered definition rather
than
$
\langle n_i(\mathbf r)n_j(\mathbf r')\rangle/n^2
$
to avoid disconnected lower-order contributions, particularly in
the same-species channel~\cite{naraschewski_spatial_1999}.
For a condensate, the macroscopic occupation of the zero-momentum
mode contributes directly to the correlation function and must be
treated separately. We therefore decompose the field operator into
a condensate background and fluctuation field,
$
\psi_i(\mathbf r)
=
\sqrt{n_0}
+
\delta\psi_i(\mathbf r),
$
with
$
\delta\psi_i(\mathbf r)
=
\frac{1}{\sqrt V}
\sum_{\mathbf k}'
e^{i\mathbf k\cdot\mathbf r}
c_{i\mathbf k}.
$
Because the system is homogeneous,
$
\mathcal G_{ij}^{(2)}(\mathbf r,\mathbf r')
=
\mathcal G_{ij}^{(2)}(\mathbf r-\mathbf r'),
$
allowing us to set $\mathbf r'=\mathbf 0$ without loss of
generality.

Substituting the condensate-fluctuation decomposition into the
four-point function yields
\begin{widetext}
\begin{equation}
\mathcal{G}_{12}^{(2)}(\mathbf r)
=
\frac{1}{n^2}
\Big\{
n_0^2
+
2n_0\tilde n
+
n_0
\Big[
2\mathrm{Re}
\big\{
\big\langle
\delta\psi_2^\dagger(\mathbf 0)
\delta\psi_1(\mathbf r)
\big\rangle
\big\}
+
\big\langle
\delta\psi_2(\mathbf 0)
\delta\psi_1(\mathbf r)
\big\rangle
+
\big\langle
\delta\psi_2^\dagger(\mathbf 0)
\delta\psi_1^\dagger(\mathbf r)
\big\rangle
\Big]
+
\big\langle
\delta\psi_2^\dagger(\mathbf 0)
\delta\psi_1^\dagger(\mathbf r)
\delta\psi_2(\mathbf 0)
\delta\psi_1(\mathbf r)
\big\rangle
\Big\}.
\end{equation}
\end{widetext}
The term $n_0^2$ represents the uniform condensate
background, while the terms proportional to $n_0$ describe the
interference between the condensate and fluctuation fields. The final
term is a four-point fluctuation contribution that contains the genuine
two-body correlations beyond the mean-field approximation.
The depletion density is obtained from the local fluctuation
correlator,
$
\tilde n_i
=
\langle
\delta\psi_i^\dagger(\mathbf 0)
\delta\psi_i(\mathbf 0)
\rangle
=
\frac{1}{V}
\sum_{\mathbf k}'
n_{i\mathbf k},
$
with an analogous definition for the anomalous average
$\tilde m_i$. Expressing the fluctuation fields in momentum space
then gives
\begin{equation}
\begin{aligned}
    \mathcal G_{12}^{(2)}(\mathbf r)
=
\frac{1}{n^2}
&\Big\{
n_0^2
+
2n_0
\Big[
\tilde n
+
\frac{1}{V}
\sum_{\mathbf k}'
e^{i\mathbf k\cdot\mathbf r}
\big(
n_{12,\mathbf k}
+
m_{12,\mathbf k}
\big)
\Big]
\\&+
\big\langle
\delta\psi_2^\dagger(\mathbf 0)
\delta\psi_1^\dagger(\mathbf r)
\delta\psi_2(\mathbf 0)
\delta\psi_1(\mathbf r)
\big\rangle
\Big\}.
\end{aligned}
\end{equation}
The remaining four-point function can be evaluated exactly using
Wick's theorem \cite{fetter_quantum_1971}. 
For the Gaussian theories considered here, it
reduces to
\begin{align} \label{eq:generic-fluctuation-correlation}
\big\langle
\delta\psi_2^\dagger(\mathbf 0)
\delta\psi_1^\dagger(\mathbf r)
\delta\psi_2(\mathbf 0)
\delta\psi_1(\mathbf r)
\big\rangle
&= \tilde n^2 \\
+ \frac{1}{V}
\Big|
\sum_{\mathbf k}'
e^{i\mathbf k\cdot\mathbf r}
n_{12,\mathbf k}
\Big|^2
&+ \frac{1}{V}
\Big|
\sum_{\mathbf k}'
e^{i\mathbf k\cdot\mathbf r}
m_{12,\mathbf k}
\Big|^2.
\nonumber 
\end{align}
We therefore evaluate the position-space correlation functions from the
normal and anomalous momentum-space correlators defined in
Eqs.~\eqref{eq:generic-correlation-coefficient-1}-\eqref{eq:generic-correlation-coefficient-4}. 
For Bogoliubov theory, we neglect the fluctuation contribution in
Eq.~\eqref{eq:generic-fluctuation-correlation}, since it enters only at
second order in the condensate depletion.

\subsection{Numerical results and discussion}

We now compare the predictions of the two theories for the
interspecies correlation functions introduced above.
These quantities provide the most direct probe of the physical
differences between bosonic pairing and Bogoliubov theories,
since they are sensitive to the treatment of the interspecies
attraction and, in particular, to the resulting structure of the
collective excitation spectrum.
The comparison therefore isolates the impact of the pairing
field $\Delta_0$ and the associated gapped spectrum on
observable correlation properties.

\begin{figure}[htbp]
    \centering
    \includegraphics[width=\columnwidth]{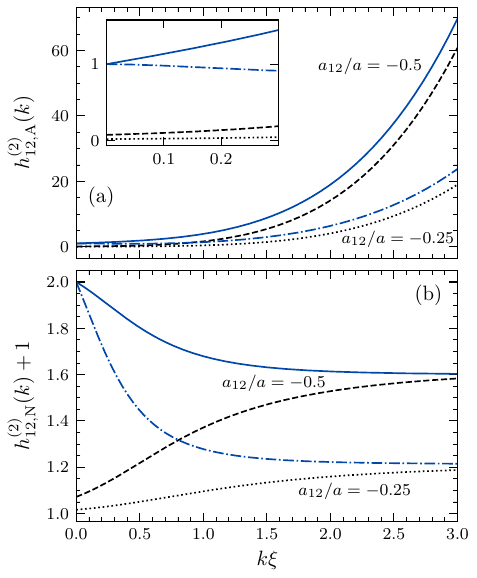}
\caption{ Momentum-space interspecies correlations for a symmetric 
two-component bosonic gas at fixed density $na^3 = 0.5\times10^{-3}$ 
in the mean-field-stable regime. 
(a) Normalized anomalous correlation $h^{(2)}_{12,A}(k)$. 
(b) Normalized normal correlation $h^{(2)}_{12,N}(k)+1$. Blue curves 
denote bosonic pairing theory, and black curves denote Bogoliubov theory. 
For bosonic pairing theory, the solid and dashed curves correspond 
to $a_{12}/a=-0.5$ and $a_{12}/a=-0.25$, respectively. For Bogoliubov 
theory, the dash-dotted and dotted curves correspond to $a_{12}/a=-0.5$ 
and $a_{12}/a=-0.25$, respectively. 
The inset in panel (a) shows the low-momentum limit. 
}
    \label{fig:corr-k}
\end{figure}

Consider Fig.~\ref{fig:corr-k}, which shows the anomalous and normal
interspecies momentum-space correlations for an equal-mass,
equal-density two-component bosonic gas, obtained from
Eqs.~\eqref{eq:norm-corr-k} and~\eqref{eq:anom-corr-k}. 
Momentum is measured in units of the single-species healing length,
$\xi=\hbar/\sqrt{2mgn}$.
The largest differences between the two approaches appear at low
momenta, where the correlations are most sensitive to the underlying
collective excitation spectrum.
At larger momenta, the predictions of the two theories converge.
We find analytically that
$
h_{12,\mathrm{N}}^{(2)}(\mathbf{k})
\underset{k\to\infty}{\longrightarrow}
\big(
\frac{B_+^2-B_-^2}{B_+^2+B_-^2}
\big)^2
$
and
$
h_{12,\mathrm{A}}^{(2)}(\mathbf{k})
\underset{k\to\infty}{\simeq}
\big[
\frac{2\epsilon_\mathbf{k}(B_- - B_+)}
{B_+^2+B_-^2}
\big]^2,
$
independently of the particular quadratic approximation employed.
For Bogoliubov theory, the anomalous correlation grows as $k^4$,
consistent with the behavior found previously for a single-component
Bose gas~\cite{butera_position-_2021}, while the normal correlation approaches
a constant value; bosonic pairing theory yields the same qualitative
ultraviolet structure.
This convergence is not accidental: if the pairing field is
approximated by its condensate contribution,
$\Delta_0 \approx -g_{12}n$. The bosonic-pairing expressions reduce
to the same ultraviolet limits as those of Bogoliubov theory.
The residual differences therefore originate from the
self-consistent anomalous average
$\tilde m_{12}$ [Eq.~\eqref{eq:BP-anom-av}], since the
kinetic-energy term dominates the quasiparticle spectrum at large
momenta and renders the details of the low-energy collective modes
increasingly unimportant.

The infrared behavior is qualitatively different between the two theories.
A $k\to0$ expansion yields
$
h^{(2)}_{12,N}(\mathbf{0})
=
\Big(
\frac{\sqrt{B_+}-\sqrt{B_-}}
     {\sqrt{B_+}+\sqrt{B_-}}
\Big)^2,
$
which evaluates to
\begin{equation}
h^{(2)}_{12,N}(\mathbf{0}) =
\begin{cases}
\displaystyle
\frac{\big(\sqrt{a^2-a_{12}^2}-a\big)^2}
     {a_{12}^2},
\quad \text{Bogoliubov,} \\[3ex]
1, 
\quad \text{bosonic pairing,}
\end{cases}
\label{eq:ir-universal}
\end{equation}
a striking contrast between the two theories.
For instance, Bogoliubov theory gives
$h^{(2),\mathrm{Bog}}_{12,\mathrm N}(\mathbf 0)\approx 0.07$
when $a_{12}/a=-0.5$, only a modest enhancement over the
uncorrelated baseline [Fig.~\ref{fig:corr-k}(b)], as expected
in a dilute gas where interactions act as a small perturbation.
In bosonic pairing theory, by contrast,
$h^{(2)}_{12,N}(\mathbf{0})$ approaches unity.
The origin lies in the excitation spectrum:
while the spin branch remains gapless in both theories, the
density branch is gapped in bosonic pairing theory, so
$\lim_{k\to0}v_{-,\mathbf k}^{2}\to\infty$
while
$\lim_{k\to0}v_{+,\mathbf k}^{2}$
approaches a finite constant.
The infrared singularity is therefore carried entirely by the spin
mode, causing the gapless branch to dominate the normalized
correlation functions and producing a result corresponding to a
doubling of the conventional Bogoliubov value.
The anomalous correlations obey the same infrared limit,
$
h^{(2)}_{12,A}(\mathbf 0)
=
h^{(2)}_{12,N}(\mathbf 0),
$
in both theories. 
For instance, this follows from the $k\to0$ expansion of
$
v_{s\mathbf k}^{2}
-
u_{s\mathbf k}v_{s\mathbf k},
$
using Eqs.~\eqref{eq:generic-usq-vsq}
and~\eqref{eq:generic-uv}, yielding the numerically verified
identity
$
h^{(2),\mathrm{BP}}_{12,A}(\mathbf 0)
=
h^{(2),\mathrm{BP}}_{12,N}(\mathbf 0)
=
1.
$

Importantly, we find that the differences between the two theories are not
confined to asymptotically small momenta.
As shown in Fig.~\ref{fig:corr-k}, significant deviations persist
up to momenta of the order of the single-species healing length,
corresponding to $k \xi\approx0.22$ in the figure, suggesting
that the predicted correlations should be experimentally observable,
particularly in light of recent measurements of momentum-space
correlations in single-species Bogoliubov
gases~\cite{tenart_observation_2021,bureik_suppression_2025}.
The anomalous correlations exhibit a shallow minimum at intermediate
momenta, marking the crossover from interaction-dominated collective
behavior to the kinetic-energy-dominated regime, beyond which the
correlations gradually approach their ultraviolet asymptotes.
The same long-wavelength limits are obtained for a weaker
attraction, $a_{12}/a=-0.25$, confirming that the enhancement is a
robust consequence of the underlying pairing physics throughout the
attractive mean-field-stable regime rather than a fine-tuned feature
of the model.

Pairing theory gives a distinct bosonic bunching mechanism that
emerges below the condensation temperature and is rooted in
pairing-induced collective correlations.
Unlike the conventional Hanbury Brown-Twiss effect, where bunching
originates from thermal occupation
statistics~\cite{hanbury_brown_correlation_1956,cayla_hanbury_2020,schellekens_hanbury_2005},
the enhancement found here survives in the low-temperature condensed
regime.
Nor is it a weak-interaction effect: Bogoliubov theory demonstrates
that small interaction parameters modify the correlations only
quantitatively, whereas the pronounced enhancement in bosonic
pairing theory reflects the fundamentally different infrared
structure produced by the gapped density mode.
The normal correlations provide the clearest distinction between the
two descriptions, with the central role played by the absence of the
interspecies normal term in the effective Hamiltonian of
Eq.~\eqref{eq:BP-BdG-matrix}.
An experimental observation of
$h^{(2)}_{12,\mathrm N}(\mathbf{0})+1 \approx 2$
would be difficult
to reconcile with conventional Bogoliubov theory and would instead
constitute strong evidence for the pairing physics captured by
bosonic pairing theory.

\begin{figure}[htbp]
    \centering
    \includegraphics[width=\columnwidth]{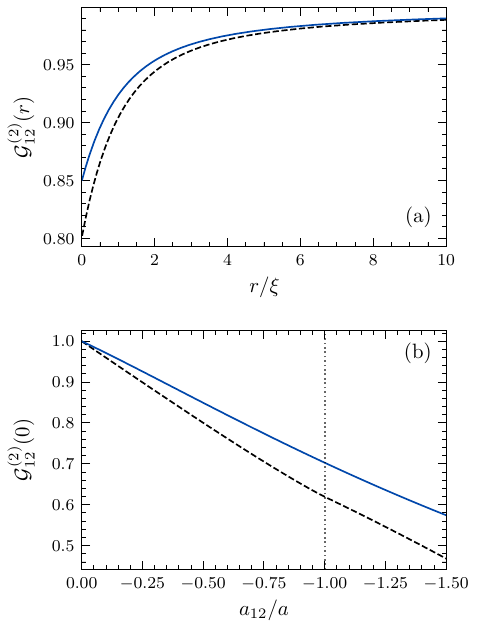}
\caption{ Ultraviolet-renormalized interspecies two-body correlation 
function $G^{(2)}_{12}(r)$ for a symmetric two-component bosonic gas. 
(a) Real-space dependence at fixed density $na^3 = 0.5\times10^{-3}$ 
and interaction ratio $a_{12}/a=-0.5$, corresponding to the attractive
mean-field-stable regime. Both theories approach the uncorrelated limit 
at large separations but differ at short distances. 
(b) Short-distance limit $G^{(2)}_{12}(0)$ across the gas-to-droplet 
crossover. The vertical dotted line marks the mean-field instability, 
$a_{12}/a=-1$, which coincides with the droplet transition. 
Blue solid curves denote bosonic pairing theory, and black dashed curves
denote Bogoliubov theory. 
In the droplet regime, the Bogoliubov result is evaluated using Petrov's 
prescription for discarding the unstable density branch.}
    \label{fig:corr-r}
\end{figure}

We now turn to the real-space correlations shown in
Fig.~\ref{fig:corr-r}, which provide direct information about the
relative probability of finding particles of different species
separated by a distance $r$.
The evaluation of $\mathcal{G}_{12}^{(2)}(\mathbf r)$ requires
Fourier transforming quantities involving $m_{12,\mathbf k}$, which
are ultraviolet divergent for contact interactions. We therefore
employ the renormalization procedure of
Eq.~\eqref{eq:renormalization}.
The short-distance limit follows analytically from the Hellmann-Feynman theorem, 
which can be written in the present case as
$ 
\mathcal{G}^{(2)}_{12}(\mathbf{0}) = \frac{1}{n^2V} \big( 
\big\langle\frac{\partial \mathcal{E}_\mathrm{mf}}
{\partial g_{12}}\big\rangle 
+ \big\langle\frac{\partial \mathcal{E}_{\mathrm{LHY}}}
{\partial g_{12}}\big\rangle \big), 
$
following Ref.~\cite{gangardt_stability_2003}.
For Bogoliubov theory, this yields
\begin{equation}
\mathcal{G}_{12}^{(2)}(\mathbf{0}) -1
= \frac{4}{3\pi^2 n}
\Big(\frac{m}{\hbar^2}\Big)^{3/2}
\big[(g_+ n)^{3/2} - (g_- n)^{3/2}\big],
\end{equation}
which is negative, indicating a suppression of short-range
interspecies correlations, consistent with the dilute weak-coupling
limit of Ref.~\cite{cherny_short-range_2000}. This expression is in
excellent agreement with the Bogoliubov theory result shown in
Fig.~\ref{fig:corr-r}(b), obtained by Fourier transforming the
momentum-space correlations.
Within Bogoliubov theory, the dominant contribution arises from
the antisymmetric branch $E_{-,\mathbf k}$, whereas the contribution
from the softening symmetric branch becomes progressively less
important as the system approaches the mean-field instability at
$a_{12}/a=-1$. At large separations, both theories recover
$\mathcal{G}_{12}^{(2)}(\mathbf r)\to1$ as $r\to\infty$, reflecting
the short-range nature of the contact interaction.

The most significant differences emerge in the short-distance
regime shown in Fig.~\ref{fig:corr-r}(b).
As the interaction strength approaches the droplet transition, the
two theories diverge increasingly: while both exhibit a suppression
of short-range correlations, bosonic pairing theory consistently
predicts a weaker antibunching-like effect than Bogoliubov theory.
The growing discrepancy mirrors the enhanced role of pairing
correlations near the droplet regime and provides an additional
observable for distinguishing the two frameworks, particularly given
recent experimental advances enabling direct in situ measurements of
spatial correlation functions in ultracold Bose
gases~\cite{yao_measuring_2025,de_jongh_quantum_2025}.

\section{Structure Factor}
\label{sec:sf}

The correlation signatures identified in the previous section
are most pronounced in the long-wavelength regime, where the
gapped density branch of bosonic pairing theory produces
qualitatively different behavior from Bogoliubov theory.
The static structure factor probes this same regime directly
and is experimentally accessible via Bragg spectroscopy ~\cite{stenger_bragg_1999},
making it a natural complement to the correlation measurements
discussed above.

We define the density-normalized species-resolved structure factors as
\begin{equation}
S_{ij}(\mathbf q)
=
\frac{1}{N}
\left\langle
\rho_{i\mathbf q}\rho_{j,-\mathbf q}
\right\rangle,
\qquad
\mathbf q \neq \mathbf 0,
\end{equation}
where
$
\rho_{i\mathbf q}
=
\sum_{\mathbf k}
c^\dagger_{i,\mathbf k+\mathbf q}
c_{i\mathbf k}
$
is the density operator. To leading order in
fluctuations,
$
\rho_{i\mathbf q}
\simeq
\sqrt{N_0}
(
c^\dagger_{i\mathbf q}
+
c_{i,-\mathbf q}
).
$
Expressing
$c_{i\mathbf q}$ in terms of the Bogoliubov quasiparticles yields
$
S_{ij}(\mathbf q)
=
\frac{n_0}{2n}
\big[
(
u_{+,\mathbf q}
+
v_{+,\mathbf q}
)^2
\pm
(
u_{-,\mathbf q}
+
v_{-,\mathbf q}
)^2
\big],
$
where the $(+)$ sign corresponds to $i=j$ and the $(-)$ sign to
$i\neq j$. Using Eqs.~\eqref{eq:generic-usq-vsq} and
\eqref{eq:generic-uv}, we obtain
$
(
u_{s\mathbf q}
+
v_{s\mathbf q}
)^2
=
(A_{s\mathbf q}-B_s)/E_{s\mathbf q},
$
giving the general expression
\begin{equation}
S_{ij}(\mathbf q)
=
\frac{n_0}{2n}
\Big(
\frac{A_{+,\mathbf q}-B_+}{E_{+,\mathbf q}}
\pm
\frac{A_{-,\mathbf q}-B_-}{E_{-,\mathbf q}}
\Big),
\end{equation}
where the $(+)$ sign corresponds to the intraspecies structure
factor and the $(-)$ sign to the interspecies structure factor.
The convolution terms arising solely from noncondensate
fluctuations are neglected in both theories, since they are beyond
the order retained in the $na^3$ expansion within the quadratic
Bogoliubov framework. The corresponding branch-resolved structure
factors are
$
S_s(\mathbf q)
=
\frac{n_0}{n}
\frac{A_{s\mathbf q}-B_s}{E_{s\mathbf q}}.
$

\begin{figure}[htbp]
    \centering
    \includegraphics[width=\columnwidth]{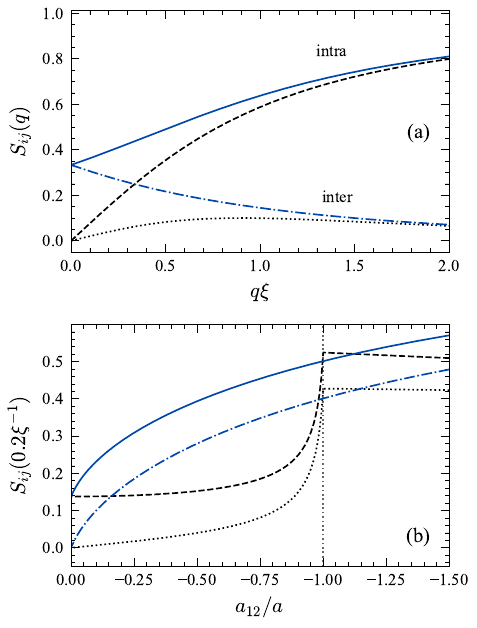}
    \caption{ Species-resolved static structure factors for a symmetric 
    two-component bosonic gas. 
    (a) Intraspecies and interspecies structure factors as functions of 
    momentum at fixed density $na^3 = 0.5\times10^{-3}$ and interaction 
    ratio $a_{12}/a=-0.5$, corresponding to the attractive mean-field-stable regime. 
    (b) Intraspecies and interspecies structure factors evaluated at 
    $q=0.2 \xi^{-1}$ across the gas-to-droplet crossover. 
    The blue solid and blue dash-dotted curves denote the intraspecies 
    and interspecies structure factors in bosonic pairing theory, 
    respectively. The black dashed and black dotted curves denote the 
    corresponding Bogoliubov results. 
    For $a_{12}<-a$, the Bogoliubov result is evaluated using Petrov's 
    prescription for discarding the unstable density branch. }
    \label{fig:sq-ij}
\end{figure}

\begin{figure}[htbp]
    \centering
    \includegraphics[width=\columnwidth]{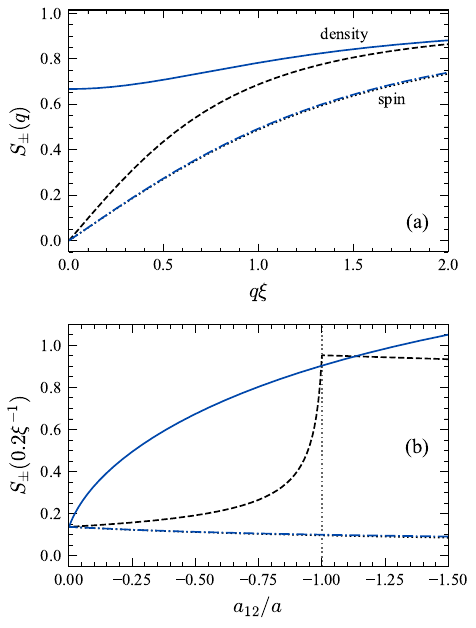}
    \caption{ Density- and spin-mode static structure factors for a 
    symmetric two-component bosonic gas.
    (a) Branch-resolved structure factors as functions of momentum at 
    fixed density $na^3 = 0.5\times10^{-3}$ and interaction ratio 
    $a_{12}/a=-0.5$, corresponding to the attractive mean-field-stable regime. 
    (b) Density and spin structure factors evaluated at $q=0.2 \xi^{-1}$ 
    across the gas-to-droplet crossover. 
    The blue solid and blue dash-dotted curves denote the density and 
    spin branches in bosonic pairing theory, respectively. The black 
    dashed and black dotted curves denote the corresponding Bogoliubov results. }
    \label{fig:sq-pm}
\end{figure}

In Fig.~\ref{fig:sq-ij}, we plot the intraspecies and interspecies
structure factors for both theories in the mean-field stable phase.
The principal difference is that, whereas in Bogoliubov theory both
branches vanish as $q\to 0$, the structure factor in bosonic pairing
theory approaches a finite positive value. The Bogoliubov result is
well established, since both density and spin modes are gapless and
satisfy the Hugenholtz-Pines
condition~\cite{sun_bogoliubov_2010}.
In bosonic pairing theory, the gap in the density mode contributes
to both the intraspecies and interspecies structure factors. For
Bogoliubov theory, 
$
S_{12}(\mathbf{q}) = \frac{n_0}{2n} \big(
\frac{\epsilon_\mathbf{q}}{E_{+,\mathbf{q}}} -
\frac{\epsilon_{\mathbf{q}}}{E_{-,\mathbf{q}}} \big),
$ 
whereas for bosonic pairing theory 
$
S_{12}(\mathbf{q}) = \frac{n_0}{2n} \big(
\frac{\epsilon_\mathbf{q}+2\Delta_0}{E_{+,\mathbf{q}}} -
\frac{\epsilon_{\mathbf{q}}}{E_{-,\mathbf{q}}} \big).
$
While $\lim_{q \to 0} S_{ij}(\mathbf{q})=0$ for a theory with both
branches gapless, bosonic pairing instead gives
\begin{equation}
     S_{ij}(\mathbf{q}) \underset{q\to 0}{\simeq}
     \frac{n_0\Delta_0}{nE_{+,\mathbf{0}}} \rightarrow
     \frac{n_0}{2n}\sqrt{\frac{\Delta_0}{g n_0}} \simeq \frac{1}{2}\sqrt{\Big| \frac{a_{12}}{a} \Big|},
\end{equation}
valid for both the interspecies and intraspecies branches.
The effects are most pronounced in the density mode $S_{+}(\mathbf
q)$, where constructive interference gives $S_{+}(\mathbf{0}) =
2S_{12}(\mathbf{0})$, as shown in Fig.~\ref{fig:sq-pm}(a). The
density branch is experimentally accessible through Bragg-type
spectroscopy, and spin-density mode separation has been demonstrated
in related
systems~\cite{sun_bogoliubov_2010,piekarski_spin_2025}.
The differences remain visible up to relatively large momenta; at
$q = 0.2/\xi$ the contrast is largest near $a_{12}/a =
-0.5$, where $S^{\text{Bog}}_{+}/S^{\text{BP}}_{+} \approx 3$, 
suggesting that Bragg
spectroscopy can provide a viable discriminant between the two
theories~\cite{stenger_bragg_1999,shammass_phonon_2012,piekarski_spin_2025}.
We also observe behavior analogous to that of Sec.~\ref{sec:corr}:
in the droplet regime, removing the unstable branch drives the
Bogoliubov result toward the bosonic pairing prediction.

As with the correlation functions, the two theories converge in the
ultraviolet. The interspecies structure factor vanishes in both,
$
S_{12}(\mathbf{q}) \underset{q\to\infty}{\simeq}
\frac{n_0\Delta_0}{n\epsilon_{\mathbf{q}}}
$ 
and 
$
S_{12}(\mathbf{q})
\simeq -\frac{g_{12}n_0}{\epsilon_{\mathbf{q}}}
$ 
for bosonic pairing and Bogoliubov theory respectively, 
while the intraspecies factor approaches $n_0/n$ in both. 
The remaining difference is traceable to
the variational determination of $n_0$ and the omission of the
interspecies normal terms in Eq.~\eqref{eq:BP-BdG-matrix}. Consistent
with this, the spin-mode structure factor is the only branch for which
the two theories remain nearly indistinguishable across interaction
strengths and over the full range of $q$, with deviations below $5\%$.
This further supports the utility of the approximation
$\tilde{m}_{12}=0$ used in Sec.~\ref{sec:pairing} to motivate the
similarities between the two theories, and highlights the restriction
to the pairing channel in Eq.~\eqref{eq:BP-BdG-matrix} as the
essential distinction of the bosonic pairing theory.

Considering now the dynamical structure factor, one immediately notices
that bosonic pairing theory violates the $f$-sum
rule~\cite{fetter_quantum_1971,pitaevskii_bose-einstein_2016,leggett_bose-einstein_2001}.
We define the first moment of the dynamic structure factor as
\begin{equation}
I_{ij}(\mathbf q)
=
\int_0^\infty d\omega\hbar\omega
S_{ij}(\mathbf q,\omega),
\end{equation}
where, for discrete quasiparticle excitations with frequencies
$\omega_{s\mathbf q}$, the dynamic structure factor is
$
S_{ij}(\mathbf q,\omega)
=
\frac{1}{2N}
\sum_e
\langle 0|
\rho_{i\mathbf q}
|e\rangle
\langle e|
\rho_{j,-\mathbf q}
|0\rangle
\delta(\omega-\omega_{e0}),
$
with $\delta(x)$ the Dirac delta function and $e$ labeling the
excited states. For a continuum system with velocity-independent
interactions and conserved particle number, the $f$-sum rule requires
this moment to equal the free-particle kinetic energy,
$
I_{ij}(\mathbf q)
=
\frac{1}{2N}
\big\langle
\big[
\rho_{i,-\mathbf q},
[H,\rho_{j\mathbf q}]
\big]
\big\rangle
= \epsilon_{\mathbf q} \delta_{ij}.
$
Evaluating the first moment within the quadratic approximation gives
$
I_{ij}(\mathbf q)
=
\frac{n_0}{2n}
\big[
(A_{+,\mathbf q}-B_+)
\pm
(A_{-,\mathbf q}-B_-)
\big],
$
where the $(+)$ and $(-)$ signs correspond to $i=j$ and $i\neq j$,
respectively. For bosonic pairing theory,
\begin{equation}
I_{11}(\mathbf q)-\epsilon_{\mathbf q}
=
\frac{n_0}{n}\Delta_0
\simeq
\Delta_0
\neq0,
\label{eq:SF_BP_density_first_moment}
\end{equation}
demonstrating a violation of the $f$-sum rule. This violation is
directly tied to the gapped density branch and its associated
structure factor. It is worth noting that the highly successful
Gross-Pitaevskii approximation also violates the $f$-sum
rule~\cite{leggett_bose-einstein_2001}; thus, the violation of an
exact relation is not necessarily an indication of a poor
approximation.

\section{Conclusion}

We have presented an operator formulation of bosonic pairing theory 
and systematically compared its predictions with those of Bogoliubov 
theory for a homogeneous symmetric Bose mixture across the gas-to-droplet 
crossover at fixed particle density.
This choice follows the experimental pathway to droplet formation and
establishes a common framework for describing both the mean-field-stable 
and droplet regimes.
By identifying the Hamiltonian-level assumptions underlying the
bosonic pairing approximation, we showed that it restricts the
attractive interspecies interaction to the zero-center-of-mass
pairing channel, thereby omitting the normal interspecies
contributions retained in Bogoliubov theory.
Despite this truncation, the variationally minimized grand potential
remains lower in bosonic pairing theory throughout the
interspecies-attractive region, confirming that the pairing ansatz
captures interspecies correlations beyond Bogoliubov theory, while both
theories recover the single-species limit as $a_{12}\to0$.

The two approaches make qualitatively different predictions for 
experimentally accessible observables.
The most pronounced signature is the strong enhancement of the
long-wavelength interspecies normal correlation,
$h^{(2)}_{12,\mathrm{N}}(\mathbf{0})+1\approx 2$ in bosonic pairing
theory versus $h^{(2)}_{12,\mathrm{N}}(\mathbf{0})+1\approx 1$ in
Bogoliubov theory.
This doubling originates from the infrared dominance of the gapless 
spin mode once the density branch becomes gapped. The singular infrared 
weight is then carried entirely by a single gapless branch, causing the 
normalized normal correlation to saturate at unity rather than the
much smaller value predicted by Bogoliubov theory.
The effect persists over a broad momentum window up to $k\xi\approx
0.2$ and is robust throughout the attractive mean-field-stable
regime, rather than being a fine-tuned feature near the droplet
transition.
In position space, the interspecies
correlation $\mathcal{G}^{(2)}_{12}(\mathbf{r})$ approaches unity at
large separations in both theories, consistent with the short-range
nature of the contact interaction.
At short distances, however, bosonic pairing theory consistently
predicts weaker antibunching, with the discrepancy incrasing
monotonically as the system approaches the droplet transition.

The static structure factor reinforces this picture.
The gapped density branch produces a finite low-momentum limit
$S_{ij}(\mathbf{q})\to(n_0/2n)\sqrt{\Delta_0/gn_0}$ as $q\to0$ in bosonic
pairing theory, whereas both branch-resolved structure factors vanish
linearly with $q$ in Bogoliubov theory.
The density-mode structure factor $S_+(\mathbf{q})$ provides the most direct
experimental probe, exhibiting a contrast ratio
$S^{\mathrm{Bog}}_+/S^{\mathrm{BP}}_+\approx3$ already at $q =
0.2/\xi$, well within the reach of Bragg spectroscopy.
By contrast, the spin-mode structure factor is nearly identical in
both theories across the entire interaction range, further confirming 
that the density channel carries the dominant signatures of pairing.
We also note that the static pairing-field approximation underlying
bosonic pairing theory leads to a violation of the $f$-sum rule in
the density channel.

Taken together, our results identify low-momentum interspecies
correlation measurements and Bragg spectroscopy across the
gas-to-droplet crossover as direct experimental tests of bosonic
pairing. The predicted signatures are sufficiently pronounced and 
robust to be clearly distinguished from the predictions of 
conventional Bogoliubov theory.
Beyond the specific system studied here, our comparison of quadratic 
theories at the level of their Hamiltonian approximations and experimentally 
accessible correlation functions provides a general framework for 
assessing alternative microscopic descriptions of quantum droplets 
and other multicomponent Bose systems.

\begin{acknowledgments}
M.I. acknowledges support from the U.S. Air Force Office of Scientific
Research (AFOSR) under Grant No.~FA8655-24-1-7391.
\end{acknowledgments}

\bibliography{bibliography}


\end{document}